\documentclass[usenatbib,usegraphicx]{mn2e}
\usepackage{amsmath}

\def\nodata{...}
\def\apj{ApJ}
\def\apjl{ApJL}
\def\apjs{ApJS}
\def\aap{A\&A}
\def\aaps{A\&AS}
\def\mnras{MNRAS}
\def\aj{AJ}
\def\araa{ARA\&A}
\def\nat{Nature}
\def\sovast{Soviet Ast.}
\def\apss{ApSS}
\def\cm{\textrm{cm}}

\def\sec{\textrm{s}}

\def\erg{\textrm{erg}}
\def\kpc{\textrm{kpc}}
\def\pc{\textrm{pc}}
\def\Mpc{\textrm{Mpc}}

\def\Kelv{\textrm{K}}
\def\Jy{\textrm{Jy}}

\def\kms{\textrm{km}~\textrm{s}^{-1}}

\def\phps{\textrm{ph}~\textrm{s}^{-1}}
\def\gcm2{\textrm{g}~\textrm{cm}^{-2}}
\def\eV{\textrm{eV}}

\def\Hz{\textrm{Hz}}
\def\kHz{\textrm{kHz}}
\def\MHz{\textrm{MHz}}
\def\GHz{\textrm{GHz}}

\def\yr{\textrm{yr}}
\def\Myr{\textrm{Myr}}

\def\muGauss{\mu\textrm{G}}

\def\Msun{\textrm{M}_{\sun}}
\def\Lsun{\textrm{L}_{\sun}}
\newcommand{\mean}[1]{\ensuremath{\langle #1 \rangle}}
\newcommand{\mcc}[1]{\multicolumn{2}{c}{#1}}
\def\bfnop{}

\title[Interpreting Starbursts at Low Frequencies]{Interpreting the Low Frequency Radio Spectra of Starburst Galaxies: A Pudding of Str\"omgren Spheres}
\author[Lacki]{Brian C. Lacki$^{1,2}$\\$^1$Jansky Fellow of the National Radio Astronomy Observatory\\$^2$Institute for Advanced Study, Einstein Drive, Princeton, NJ 08540, USA, brianlacki@ias.edu}

\date{Draft Version}

\begin{document}

\maketitle

\begin{abstract}
The low frequency radio emission of starburst galaxies is informative, but it can be absorbed in several ways.  Most importantly, starburst galaxies are home to many H II regions, whose free-free absorption {\bfnop blocks} low frequency radio waves.  These H II regions are discrete objects, but most multiwavelength models of starbursts assume a uniform medium of ionized gas, if they include the absorption at all.  I calculate the effective absorption coefficient of H II regions in starbursts, which is ultimately a cross section times the density of H II regions.  The cross sections{\bfnop are} calculated by assuming that H II regions are Str\"omgren spheres.  The coefficient asymptotes to a constant value at low frequencies, because H II regions partially cover the starburst, and are buried part way into the starburst's synchrotron emitting material.  Considering Str\"omgren spheres around {\bfnop either OB} stars {\bfnop or} Super Star Clusters, I {\bfnop demonstrate the method by fitting to the} low frequency radio spectrum of M82.  {\bfnop I discuss implications of the results for synchrotron spectrum shape, H II region pressure, and free-free emission as a star-formation rate indicator.  However, these results are preliminary, and could be affected by systematics.  I argue that there is no volume-filling warm ionized medium in starbursts, and that H II regions may be the most important absorption process down to $\sim 10\ \MHz$.  Future data at low and high radio frequency will improve our knowledge of the ionized gas.}
\end{abstract}

\section{Introduction}
\label{sec:Introduction}
The observational prospects for low frequency studies of star-forming galaxies are {\bfnop improving}.  There is increasing interest in low frequency radio {\bfnop instruments}, due to their value in observing high redshift 21 cm lines, among other reasons.  The Giant Metrewave Radio Telescope (GMRT) is specifically designed to provide interferometric data for radio sources in the 50 MHz to 1.5 GHz range with high sensitivity\footnote{See http://www.gmrt.ncra.tifr.res.in.}.  The 74 and 333 MHz systems on the Very Large Array (VLA) provide high angular resolution images \citep{Kassim07}, and have completed a survey of the northern sky \citep{Cohen07}.  {\bfnop They are} currently being upgraded for use on the {\bfnop Karl G.} Jansky VLA.  The Low Frequency Array (LOFAR), now coming online, is a new radio telescope with long baseline interferometry capabilities, and can go all the way down to 15 MHz\footnote{At http://www.lofar.org.}.  The Leiden LOFAR Sky Surveys Project\footnote{With a home page at http://lofar.strw.leidenuniv.nl/.} will image some nearby star-forming galaxies in radio, possibly including starbursts like M82.  LOFAR will be joined by the 21 cm pathfinder experiments at frequencies above 100 MHz and possibly the Long Wavelength Array at frequencies of 10 to 88 MHz \citep[e.g.,][]{Ellingson09}.  Ultimately, the Square Kilometre Array (SKA) should be able to observe galaxies down to 70 MHz with high sensitivity\footnote{http://www.skatelescope.org.}.  

The radio spectrum of star-forming galaxies is dominated by synchrotron emission from cosmic ray (CR) electrons and positrons ($e^{\pm}$) in diffuse magnetic fields \citep{Condon92}.  There is also free-free emission, which typically comes from H II regions in a galaxy.  Both of these emission processes are associated with star-formation: cosmic rays are generated somehow by star formation (possibly through shock acceleration in supernova remnants), and H II regions surround young, massive stars that produce ionizing radiation.  Since synchrotron emission has a steeply falling spectrum (typically $S_{\nu}^{\rm synch} \propto \nu^{-0.7}$) whereas free-free emission does not ($S_{\nu}^{\rm ff} \propto \nu^{-0.1}$), the synchrotron emission dominates below about 30 GHz (e.g., \citealt{Condon92}; \citealt*{Niklas97}).  

{\bfnop While these processes are both present in nearly all star-forming galaxies, the physical conditions within star-forming galaxies can vary tremendously.  Extreme environments for star-formation can be found in starbursts (defined here as regions with star-formation rate surface densities $\Sigma_{\rm SFR} \ga 1\ \Msun\ \yr^{-1}\ \kpc^{-2}$).  Examples of starburst regions include the Galactic Centre CMZ and those found in the galaxies NGC 253, M82, and Arp 220.  Average densities and pressures in these regions can be hundreds of times greater than in the present-day Milky Way, altering the structure of the ISM.}

{\bfnop A} wealth of information {\bfnop on how these environments affect the CRs and ionized gas} is available at MHz frequencies.  In particular, different cooling processes may set the CR $e^{\pm}$ lifetime at different energies.  At low frequencies, bremsstrahlung, with an energy-independent loss time, and ionization, which is most effective at low energies, become more important than synchrotron and Inverse Compton cooling, which grow stronger at high energies (\citealt{Hummel91,Thompson06,Murphy09}; \citealt*{Lacki10-FRC1}).  Any {\bfnop CR} escape, whether diffusive or advective, also becomes more important relative to synchrotron at low frequencies.  Therefore, the synchrotron radio spectrum should flatten at low frequencies and steepen at high frequencies.  Indeed, this behaviour is often seen in the radio spectra of starburst galaxies \citep[e.g.,][]{Clemens08,Williams10,Leroy11}.  The detailed radio spectra {\bfnop are} useful in constructing models of the cosmic ray population, helping to constrain the poorly understood magnetic field strength (e.g., \citealt{Torres04-Arp220,Domingo05}; \citealt*{Persic08}; \citealt*{deCeaDelPozo09-M82}; \citealt{Lacki10-FRC1}; \citealt*{Rephaeli10}; \citealt{Crocker11-Wild}).  Low frequency synchrotron emission might also {\bfnop betray} the presence of a `pion bump' in the $e^{\pm}$ spectrum, in which the spectrum of secondary $e^{\pm}$ from pion decay falls off due to the kinematics of pion production in proton-proton collisions \citep{Rengarajan05}.  {\bfnop These processes are expected to be particularly important in starburst regions \citep{Lacki10-FRC1}.}

In practice, our ability to understand the low frequency radio spectrum of starbursts is limited by other processes {\bfnop that} alter the radio spectrum.  The most important is free-free absorption by {\bfnop the} ionized gas in galaxies.  The Galactic radio spectrum has a turnover at $\sim 3\ \MHz$, largely caused by free-free absorption in the diffuse Warm Ionized Medium (WIM; {\bfnop \citealt{Hoyle63};} \citealt{Alexander69,Fleishman95,Peterson02}).  Free-free absorption is even more important along lines of sight through dense H II regions, which can become optically thick even at GHz frequencies \citep[e.g.,][]{McDonald02}.  {\bfnop Thus,} free-free absorption can {\bfnop also} flatten the low frequency radio spectra of starbursts.  The spectral curvature of starburst galaxies has therefore been interpreted as free-free absorption (\citealt*{Klein88}; \citealt{Carilli96,Clemens10}).

However, there has been relatively little work on the theory of the low frequency radio spectra of starburst galaxies.  For model fitting, {\bfnop if free-free absorption is even considered at all, the typical assumption is the uniform slab model,} in which both the free-free absorption and emission come from a uniform density ionized medium pervading the synchrotron-emitting region (for examples of uniform slabs used to fit starburst radio spectra, see, e.g., \citealt{Sopp91,Condon91,Carilli96,Torres04-Arp220,Clemens10,Williams10,Adebahr12}).  The best {\bfnop current} measurements are at GHz frequencies, where the integrated free-free absorption is often expected to be small and the details of the absorption may not matter much, {\bfnop although} it has been claimed to be important at GHz frequencies in Arp 220 and other Ultraluminous Infrared Galaxies (ULIRGs) \citep{Condon91,Sopp91,Clemens08,Clemens10}.  However, from a theoretical point of view, this approximation is likely to be too simple: {\bfnop even} in the Milky Way, the free-free absorption largely comes from the {\bfnop WIM}, {\bfnop but} the free-free emission comes from compact H II regions.  

{\bfnop Unlike the Milky Way and other normal star-forming galaxies, the H II regions could actually dominate the free-free absorption within starburst regions.  Although much of the gas mass in normal galaxies is warm gas (both neutral and ionized), theories of starburst ISM suggest this is not the case in these intense regions.  Instead, it is more likely that the starburst volume is filled by either the hot ($\sim 10^8\ \Kelv$) plasma excavated out by supernovae that forms into the wind (\citealt*{Heckman90}; \citealt{Lord96}), or by the dense ($\ga 100\ \cm^{-3}$) molecular gas that makes up most of the mass (\citealt*{Thompson05}).  In the hot wind, ionizing photons escape readily without interacting with the neutral gas; in the molecular gas, they could be stopped too quickly by the enormous absorbing columns.  Another hurdle for the formation of a WIM is the extraordinary pressures within starburst regions ($\ga 10^7\ \Kelv\ \cm^{-3}$ in M82).  Observationally, in starburst regions, at least part of the free-free absorption is directly observed to come from discrete H II regions, as seen in the Galactic Centre at 74 and 333 MHz \citep{Brogan03,Nord06}, and in M82 \citep{Wills97}. Finally, the low frequency spectra of some supernova remnants in M82 do not show evidence of free-free absorption \citep{Wills97}, which is consistent with the absorption being patchy.}

The alternative is to consider {\bfnop absorption from a collection of discrete H II regions in starbursts}.  A simple version of this approach has been considered in the context of radio recombination line studies, where the radio spectrum is needed to calculate stimulated emission \citep[e.g.,][]{Anantharamaiah93,Zhao96,Rodriguez-Rico05}.  The {\bfnop usual} assumption in these models is that all of the H II regions have the same density, temperature, and radius, and are located in the midplane of a starburst disc.  Each H II region then shadows the synchrotron-emitting region behind it{\bfnop ;} with this assumption, the free-free absorption can be predicted \citep[e.g.,][]{Anantharamaiah93}.  However, the model does not work well when the H II regions are both optically thick and have a covering fraction near 100\% \citep{Anantharamaiah93}, and it does not allow for H II regions of different radii.  

In this paper, I study effects on the low frequency spectra of starbursts.  I focus on free-free absorption {\bfnop from H II regions}, the most important process, {\bfnop but I also check whether other processes are important}.  I calculate {\bfnop in Section~\ref{sec:Scenarios}} the amount of free-free absorption by assuming it comes from discrete H II regions around {\bfnop OB stars} stellar clusters in starbursts.  In essence, {\bfnop I use a `raisin pudding'-like model of starbursts: the dark, opaque Str\"omgren spheres are mixed in with a surrounding (transparent) volume-filling medium}.  My model generalizes the approach of \citet{Anantharamaiah93} to H II regions of different radii and {\bfnop lets} optically thick H II regions {\bfnop have} high covering fraction.  I make spectral fits using these model to the low frequency radio spectrum of M82 in Section~\ref{sec:M82Fit}, accounting for the possibility of spectral curvature.  I {\bfnop check the range of validity of the models by computing whether} free-free absorption from {\bfnop diffuse ionized gas}, the Razin effect, or synchrotron self-absorption, cut{\bfnop s} off the low frequency spectrum in Section~\ref{sec:StarburstRadioEnd}.  {\bfnop Some details of the calculations are presented in the Appendices.}

\section{How Populations of H II regions Absorb Radio Emission in Starbursts}
\label{sec:Scenarios}
The nonthermal continuum radio emission of starbursts are thought to pervade the entire starburst region, because the radiating cosmic rays can diffuse from their acceleration sites \citep[e.g.,][]{Torres12}.  The H II regions are intermixed in the starburst region, so in this sense the uniform slab model is correct.  {\bfnop A} model of truly uniform ionized gas assumes the free-free absorption comes from a low density, high filling factor medium, {\bfnop one with} a low turnover frequency but {\bfnop is} highly opaque below {\bfnop that} frequency.  However, a more appropriate assumption is that there is a uniform density of H II regions rather than of ionized gas.  A uniform collection of discrete H II regions is a high density, low filling factor medium, with a higher turnover frequency, but {\bfnop potentially} translucent below that frequency.

{\bfnop I show in Appendix~\ref{sec:Derivation} that the absorption from H II regions can be parametrized with an absorption coefficient,
\begin{equation}
\label{eqn:alphaEff}
\alpha_{\rm eff} \approx \int \frac{dN}{dq dV} \sigma(q) dq.
\end{equation}
In this equation, $dN/(dq dV)$ is a distribution function over some parameter(s) $q$ that gives the number of each type of H II region per unit volume, and $\sigma(q)$ is an effective absorbing cross section for that type of H II region.}

{\bfnop The simplest possible assumption is that H II regions are uniform density Str\"omgren spheres, with a radius of} 
\begin{equation}
R_S = \left(\frac{3 Q_{\rm ion}^{\star}}{4 \pi n_H^2 \alpha_B}\right)^{1/3},
\end{equation}
{\bfnop for an ionizing photon injection rate $Q_{\rm ion}^{\star}$ per H II region.}  The recombination constant $\alpha_B$ is equal to $\alpha_B = 2.56 \times 10^{-13} (T/10^4\ \Kelv)^{-0.83} \cm^3 \sec^{-1}$ \citep{Draine11-Book}.  {\bfnop I assume that the Str\"omgren spheres are completely ionized inside $R_S$ and are completely neutral outside $R_S$.  I summarize the absorption and emission properties of Str\"omgren spheres in Appendix~\ref{sec:StromgrenRadiative}.  Equation~\ref{eqn:SigmaTranslucentSphere} gives the effective cross section of a uniform density sphere with some absorptivity.  Note the free-free absorption coefficient within each Str\"omgren sphere, if it is fully ionized hydrogen plasma (with $n_e = n_H$), is}
\begin{equation}
\alpha_{\rm H II} = 0.018\ \cm^{-1} \left(\frac{n_e^2 T^{-3/2} \nu^{-2} \bar{g_{\rm ff}}}{\cm^{-6} \Kelv^{-3/2} \Hz^{-2}}\right),
\end{equation}
with the Gaunt factor $\bar{g_{\rm ff}}$ usually having a value near $\sim 10$ for $10^4\ \Kelv$ plasma at MHz frequencies \citep{Rybicki79}.\footnote{{\bfnop The exact value of the Gaunt factor I use when calculating $\alpha_{\rm eff}$ is $\bar{g_{\rm ff}} = 6.155 (\nu / \GHz)^{-0.118} (T / 10^4\ \Kelv)^{0.177}$, from \citet{Draine11-Book}.}}

The other necessary ingredient is the distribution function $dN/(dq dV)$, which depends on the number and properties of ionizing sources.  I consider Str\"omgren spheres surrounding several possible types of ionizing sources: identical sources with the same luminosity, individual stars in a realistic stellar population, or stellar clusters that either shine at a constant luminosity for some time and shut off or fade more realistically.

\subsection{The simple source scenario}
\label{sec:IndivOStars}
{\bfnop Suppose that all of the ionizing sources have the same ionizing luminosity, and the H II regions around them have the same physical properties (e.g., temperature, density).  For example, a common order of magnitude estimate for a typical OB star's ionizing photon luminosity is $Q_{\rm ion}^{\star} \approx 10^{49}\ \phps$.  

For a long-lived ($\ga 10\ \Myr$) continuous starburst, \citet{Leitherer99} finds the ionizing photon luminosity as
\begin{equation}
\label{eqn:QwSFR}
Q_{\rm ion} = 2.18 \times 10^{53} \phps \left(\frac{\rm SFR}{\Msun\ \yr^{-1}}\right),
\end{equation}
assuming Solar metallicity and a Salpeter mass function from 0.1 to $100\ \Msun$.  I divide the total ionizing photon flux of eqn.~\ref{eqn:QwSFR} by $Q_{\rm ion}^{\star}$ to get} the number of {\bfnop ionizing sources} in the galaxy:
\begin{equation}
N_{\star} = 21800 \left(\frac{\rm SFR}{\Msun\ \yr^{-1}}\right) \left(\frac{Q_{\rm ion}^{\star}}{10^{49}\ \phps}\right)^{-1}.
\end{equation}

The Str\"omgren radius around each {\bfnop source} is
\begin{equation}
R_S = 0.68\ \pc \left(\frac{Q_{\rm ion}^{\star}}{10^{49}\ \phps}\right)^{1/3} \left(\frac{n_H}{1000\ \cm^{-3}}\right)^{-2/3} {\bfnop \left(\frac{T}{10^4\ \Kelv}\right)^{0.28}}.
\end{equation}
Therefore, each H II region becomes optically thick ($\alpha_{\rm H II} R_S = 1$) when
\begin{eqnarray}
\nonumber \nu_S & = & 615\ \MHz \left(\frac{T}{10^4\ \Kelv}\right)^{\bfnop -0.61} \left(\frac{n_H}{1000\ \cm^{-3}}\right)^{2/3} \\
                & & \times \left(\frac{Q_{\rm ion}^{\star}}{10^{49}\ \phps}\right)^{1/6} \left(\frac{\bar{g_{\rm ff}}}{10}\right)^{1/2}
\end{eqnarray}

If the Str\"omgren spheres are in a uniform density medium, then all of the spheres are the same size, {\bfnop and the effective absorption coefficient is very simply $\alpha_{\rm eff} = N_{\star} \sigma / V_{\rm SB}$. At low frequencies, when the H II regions are all opaque ($\nu \la \nu_S$)}, I find
\begin{eqnarray}
\label{eqn:SimpleSourceAlpha}
\nonumber \alpha_{\rm eff} & \approx & (31\ \pc)^{-1} {\bfnop \left(\frac{\rho_{\rm SFR}}{1000\ \Msun\ \yr^{-1}\ \kpc^{-3}}\right)} \\
& & \times  \left(\frac{Q_{\rm ion}^{\star}}{10^{49}\ \sec^{-1}}\right)^{-1/3} \left(\frac{n_H}{1000\ \cm^{-3}}\right)^{-4/3} {\bfnop \left(\frac{T}{10^4\ \Kelv}\right)^{0.55}}.
\end{eqnarray}
{\bfnop For example, in the starburst M82 with $\rho_{\rm SFR} \approx 500\ \Msun\ \yr^{-1}\ \kpc^{-3}$ and average densities of $n_H \approx 400\ \cm^{-3}$, there would be $\sim 20\ \pc$ on any line of sight before hitting an H II region.}  Since the typical line of sight through M82 has length $\sim R_{\rm SB} \approx 250\ \pc$, {\bfnop about $\sim 10\%$ of the low frequency flux would be transmitted at low frequencies.  Of course, this assumes that all of the extreme UV flux ionizes gas, but some may be absorbed by dust or escape the starburst entirely.}

\begin{figure}
\centerline{\includegraphics[width=9cm]{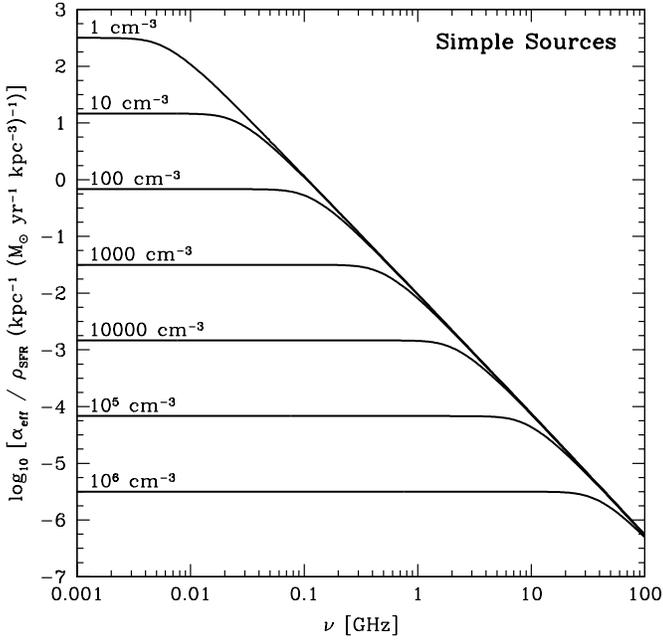}}
\caption{Effective absorption coefficient for H II regions of various densities, as normalized to a volumetric star-formation rate of $1\ \Msun\ \yr^{-1}\ \kpc^{-3}$.  This plot shows the case when {\bfnop all ionizing sources have $Q_{\rm ion}^{\star} = 10^{49}\ \sec^{-1}$}.\label{fig:AlphaEffWNuIndiv}}
\end{figure}

In Figure~\ref{fig:AlphaEffWNuIndiv}, I show the effective absorption coefficient from Str\"omgren spheres around {\bfnop simple sources} for densities $1 - 10^6\ \cm^{-3}$.  The absorption coefficients reach a plateau at low frequency, with a value that depends on hydrogen density.  At high frequencies, the absorption coefficients for different hydrogen densities all have the same value.  In this case, the cross section of {\bfnop each} Str\"omgren sphere with volume $V_S = 4/3 \pi R_S^3$ is $V_S \alpha_{\rm H II}$ (section~\ref{sec:EffectiveSigma}), which does not depend on density.

\subsection{H II regions around individual stars in a realistic stellar population}
\begin{figure}
\centerline{\includegraphics[width=9cm]{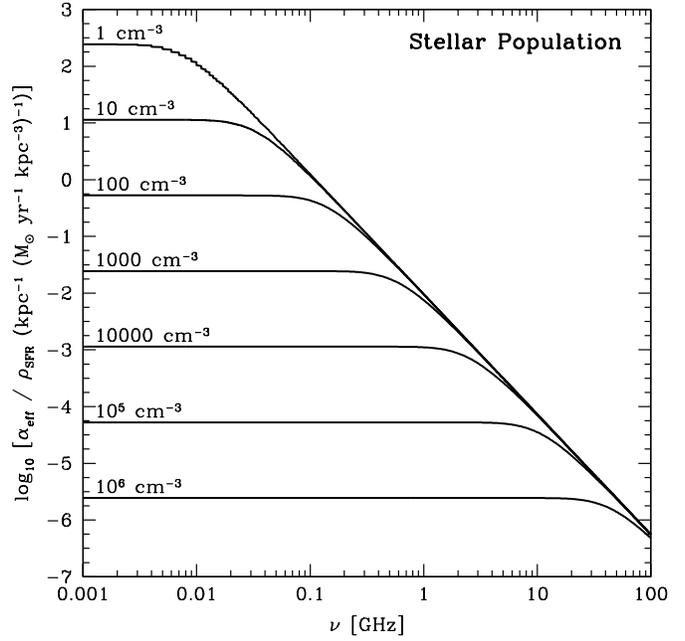}}
\caption{Effective absorption coefficient for H II regions of various densities, as normalized to a volumetric star-formation rate of $1\ \Msun\ \yr^{-1}\ \kpc^{-3}$.  This plot shows the case for a realistic population of O, B, and Wolf-Rayet stars.\label{fig:AlphaEffWNuPopIndiv}}
\end{figure}

{\bfnop Stars in real stellar populations have differing ionizing photon luminosities, which means that the Str\"omgren spheres can have differing sizes.  The distribution function of $Q_{\rm ion}^{\star}$ can be calculated by combining a model stellar population with the ionizing photon luminosity of each type of star.  To this end I ran Starburst99 models, described in detail in Appendix~\ref{sec:PopIndivDist}.

Using the distribution function from Appendix~\ref{sec:PopIndivDist}, I compute the $\alpha_{\rm eff}$ of Str\"omgren spheres of uniform density and a constant temperature as $\int dN/(dQ_{\rm ion}^{\star} dV) \sigma(Q_{\rm ion}^{\star} dQ_{\rm ion}^{\star}$.  The coefficients as functions of frequency are depicted in Figure~\ref{fig:AlphaEffWNuPopIndiv} for $T = 10^4\ \Kelv$.  At low frequencies, the coefficient asymptotes to 
\begin{eqnarray}
\nonumber \alpha_{\rm eff} & \approx & (41\ \pc)^{-1} \left(\frac{\rho_{\rm SFR}}{1000\ \Msun\ \yr^{-1}\ \kpc^{-3}}\right)\\
& & \times  \left(\frac{n_H}{1000\ \cm^{-3}}\right)^{-4/3} \left(\frac{T}{10^4\ \Kelv}\right)^{0.55}.
\end{eqnarray}
This is similar to the low frequency value in the simple source model (equation~\ref{eqn:SimpleSourceAlpha}).  The main difference with the simple sources model is at intermediate frequencies, as the H II regions start becoming transparent.  Since the Str\"omgren spheres have differing radii in this scenario, the transition from opaqueness to transparency is spread over a larger frequency range.}

\subsection{H II regions around Super Star Clusters: A simple model}
\label{sec:SSCs}
Much of the star-formation in starburst galaxies occurs in bound {\bfnop S}uper {\bfnop S}tar {\bfnop C}lusters (SSCs; e.g., \citealt{OConnell95,Melo05,Smith06}).  The SSCs have a mass function that can be described with a Schecter mass function
\begin{equation}
\label{eqn:SSCIMF}
\frac{dN}{dM_{\star}} = C M_{\star}^{-2} \exp\left(-\frac{M_{\star}}{M_c}\right),
\end{equation}
where $M_c$ is a cutoff mass {\bfnop that is} around $5 \times 10^6 \Msun$ in starbursts \citep{Meurer95,McCrady07}, and the mass function applies only above a lower mass limit $M_l$, which I take to be $1000\ \Msun$.

{\bfnop Now we must relate the stellar mass of each SSC (of a given age) with an ionizing photon luminosity.  The simplest evolution of $Q_{\rm ion}$ is that it is constant for some time $t_{\rm ion}$ and then instantly shuts off.  So} suppose that 
\begin{itemize}
\item The initial mass function of SSCs has the same form as the observed mass function.
\item The ionizing photon luminosity $Q_{\rm ion}^{\star}$ of a SSC is directly proportional to its stellar mass, {\bfnop is constant for ages up to $t_{\rm ion} = 10\ \Myr$, and is zero afterwards.}
\item The starburst has been continuously forming stars (and SSCs) at a constant rate for a time $t_{\rm burst} > t_{\rm ion}$.  
\end{itemize}

Then I can construct a mass function of {\bfnop the} SSCs {\bfnop young enough to host} ionizing stars:
\begin{equation}
\frac{dN_{\rm ion}}{dM_{\star}} = C_{\rm ion} M_{\star}^{-2} \exp\left(-\frac{M_{\star}}{M_c}\right),
\end{equation}
which is normalized so that $M_{\star}^{\rm ion} = {\rm SFR} \times t_{\rm ion} = \int_{M_l}^{\infty} C_{\rm ion} M_{\star}^{-1} \exp(-M_{\star}/M_c) dM_{\star}$.  For $M_l = 1000\ \Msun$ and $M_{c} = 5 \times 10^6\ \Msun$, $C_{\rm ion} = 0.126\ {\rm SFR} \times t_{\rm ion}$.

Under my assumptions, these ionizing photons all come from stars with ages less than $t_{\rm ion}$.  We can therefore convert the star-formation into the {\bfnop total} mass of {\bfnop all} the SSCs containing ionizing stars as {\bfnop $M_{\star}^{\rm ion} = {\rm SFR} \times t_{\rm ion}$}:
\begin{equation}
Q_{\rm ion} = 2.18 \times 10^{46} \phps \left(\frac{M_{\star}^{\rm ion}}{\Msun}\right) \left(\frac{t_{\rm ion}}{10\ \Myr}\right)^{-1}.
\end{equation}
Plugging in typical values for an SSC in a starburst, I find
\begin{align}
\nonumber R_S & = 8.8\ \pc \left(\frac{M_{\star}}{10^6\ \Msun}\right)^{1/3} \left(\frac{n_H}{1000\ \cm^{-3}}\right)^{-2/3}\\
& \times \left(\frac{t_{\rm ion}}{10\ \Myr}\right)^{-1/3} \left(\frac{T}{10^4\ \Kelv}\right)^{0.28}.
\end{align}

The absorption coefficient at low $\nu$ from H II regions is, after integrating the SSC mass function from $M_l$ to infinity, 
\begin{equation}
\alpha_{\rm eff} = \frac{3 C_{\rm ion} \pi R_0^2}{V M_0^{2/3}} \left[\frac{\exp(-M_l / M_c)}{M_l^{1/3}} - \frac{\Gamma(2/3, M_l / M_c)}{M_c^{1/3}}\right],
\end{equation}
where I take $R_S = R_0 (M / M_0)^{1/3}$.  For the case when $M_l = 1000\ \Msun$ and $M_c = 5 \times 10^6\ \Msun$, I find
\begin{align}
\nonumber \alpha_{\rm eff} & \approx (120\ \pc)^{-1} {\bfnop \left(\frac{\rho_{\rm SFR}}{1000\ \Msun\ \yr^{-1}\ \kpc^{-3}}\right)}\\
& \times  \left(\frac{n_H}{1000\ \cm^{-3}}\right)^{-4/3} \left(\frac{t_{\rm ion}}{10\ \Myr}\right)^{\bfnop 1/3} {\bfnop \left(\frac{T}{10^4\ \Kelv}\right)^{0.55}}.
\end{align}
The absorption coefficient is smaller than in the case of each OB star having its own Str\"omgren sphere.  {\bfnop The reason} can be {\bfnop understood} if we assume each cluster has $N_{\star}$ OB stars, each with the same ionizing luminosity.  Since the Str\"omgren radius {\bfnop increases} only as $N_{\star}^{1/3}$, the cross section of each H II region increases only as $N_{\star}^{2/3}$.  The number of Str\"omgren spheres instead decreases as $N_{\star}^{-1}$, meaning the effective absorption coefficient is proportional to $N_{\star}^{-1/3}$: clustered stars are not as effective at obscuration.  Essentially, clustering preserves the filling factor that is ionized, but since the {\bfnop ionized} regions are spatially correlated -- the ionized regions at the front of a large H II region already obscures the ionized volume behind it -- the covering factor is decreased.  

\begin{figure}
\centerline{\includegraphics[width=9cm]{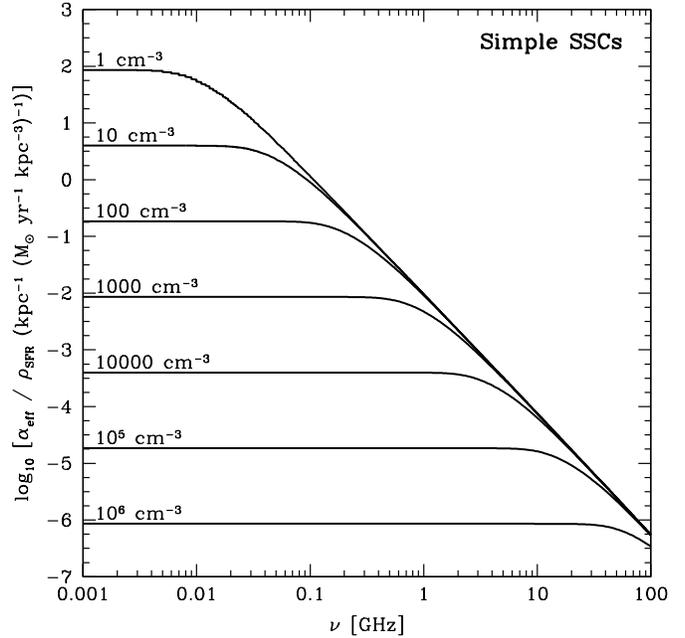}}
\caption{Effective absorption coefficient for H II regions of various densities, as normalized to a volumetric star-formation rate of $1\ \Msun\ \yr^{-1}\ \kpc^{-3}$.  This plot shows the case when stars are all within `simple' SSCs surrounded by H II regions, in which the ionizing photon luminosity of a SSC remains constant for $t_{\rm ion} = 10\ \Myr$ and then shuts off.  The low mass cutoff is $M_l = 1000\ \Msun$, and the characteristic highest mass is $M_c = 5 \times 10^6\ \Msun$. \label{fig:AlphaEffWNuClust}}
\end{figure}

I plot in Figure~\ref{fig:AlphaEffWNuClust} the effective absorption coefficients for Str\"omgren spheres around {\bfnop these `simple'} SSCs with masses above $1000\ \Msun$ and a cutoff of $5 \times 10^6\ \Msun$.  The $\alpha_{\rm eff}$ {\bfnop values} are indeed lower than in the case of individual stars, but the functional form is basically the same, with a plateau at low frequencies that depends on hydrogen density and an asymptotic form at high frequencies that is independent of hydrogen {\bfnop density}.

\subsection{H II regions around Super Star Clusters: A model that includes aging}
\label{sec:AgingSSCs}
My assumption in the previous subsection -- that SSCs behave like light bulbs, emitting ionizing photons for some time before shutting off -- is simplistic.  In fact, with {\bfnop stellar population} models {\bfnop like} Starburst99, it is possible to predict how the ionizing photon generation rate evolves for stellar populations of various ages \citep{Leitherer99}.  In principle, one can then use a known star-formation history to accurately predict the cluster mass and age distribution function, and then integrate the cross sections to get an effective absorption coefficient.

I {\bfnop now} consider the case when the star-formation rate has been constant for a duration $t_{\rm burst}$, before which it was zero.  The SSC distribution function is then assumed to have the form
\begin{equation}
\frac{dN}{dM_{\star} dt} = \frac{C_{\rm aging}}{t_{\rm burst}} M_{\star}^{-2} \exp\left(-\frac{M_{\star}}{M_c}\right)
\end{equation}
where $M_{\star}$ is the mass of stars initially formed in the cluster, and $t$ is the age of the cluster.  The SSC initial mass function then has the same form as before, in equation~\ref{eqn:SSCIMF}.  The normalization $C_{\rm aging}$ is again set by integrating over masses to get the star-formation rate, ${\rm SFR} = \int_{M_l}^{\infty} M_{\star} dN/(dM_{\star} dt) dM_{\star}$.  For my standard values of $M_l = 1000\ \Msun$ and $M_c = 5 \times 10^6\ \Msun$, I find $C_{\rm aging} = 0.126\,{\rm SFR} \times t_{\rm burst}$.

I {\bfnop compute} the effective absorption coefficient by integrating the H II region absorption cross section over different cluster masses and ages:
\begin{equation}
\alpha_{\rm eff} = \int_{M_l}^{\infty} \int_{0}^{t_{\rm burst}} \frac{dN}{dM_{\star} dt} \frac{1}{V_{\rm SB}} \sigma (Q_{\rm ion}^{\star} (M_{\star}, t), \nu) dt dM_{\star}.
\end{equation}
The starburst volume is here denoted as $V_{\rm SB}$.  The ionizing photon rate {\bfnop per unit mass} for a stellar population of age $t$ is given in \citet{Leitherer99}.  

\begin{figure}
\centerline{\includegraphics[width=9cm]{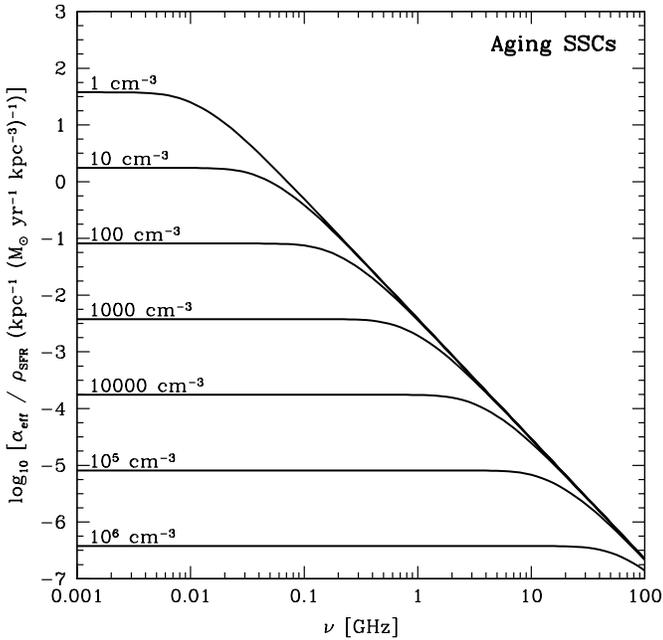}}
\caption{Effective absorption coefficient for H II regions around aging SSCs of various densities, as normalized to a volumetric star-formation rate of $1\ \Msun\ \yr^{-1}\ \kpc^{-3}$.  The starburst age is assumed to be 10 Myr.  The low mass cutoff is $M_l = 1000\ \Msun$, and the characteristic highest mass is $M_c = 5 \times 10^6\ \Msun$.\label{fig:AlphaEffWNuAC}}
\end{figure}

I show the resulting $\alpha_{\rm eff}$ for a 10 Myr old continuously forming starburst in Figure~\ref{fig:AlphaEffWNuAC}.  The opacities are lower than in the simply-modelled SSCs plotted in Figure~\ref{fig:AlphaEffWNuClust}.  {\bfnop The low frequency effective absorption coefficient is
\begin{eqnarray}
\nonumber \alpha_{\rm eff} & \approx & (260\ \pc)^{-1} \left(\frac{\rho_{\rm SFR}}{1000\ \Msun\ \yr^{-1}\ \kpc^{-3}}\right)\\
& & \times  \left(\frac{n_H}{1000\ \cm^{-3}}\right)^{-4/3} \left(\frac{T}{10^4\ \Kelv}\right)^{0.55}.
\end{eqnarray}
Note that with average values for M82, the effective absorptivity would be $\sim (150\ \pc)^{-1}$, meaning that there are roughly 1 or 2 H II regions per sightline.}

\section{Demonstration with Fits to M82's Radio Spectrum}
\label{sec:M82Fit}
{\bfnop Determining the amount of free-free absorption in starburst regions has several important applications.  The first is to accurately measure the underlying synchrotron spectrum.  The spectral index and spectral curvature constrain the lifetime and sources of GHz-emitting CR $e^{\pm}$, which may be very different at low frequencies and in the extreme environments in starburst regions \citep[e.g.,][]{Hummel91,Rengarajan05,Thompson06}.  The shape of the synchrotron spectrum is also necessary for the interpretation of radio recombination line observations.  Another motivation is to understand the properties of the ionized gas that is responsible for free-free absorption.  In particular, the thermal pressure can be calculated from the density and temperature \citep[e.g.,][]{Carilli96}.  If this thermal pressure is much smaller than the known pressure of other ISM phases in starbursts, this is evidence for nonthermal support (for example, by turbulence; c.f., \citealt{Smith06}).  Finally, the ionized gas contributes to the free-free emission.  Knowledge of the amount of free-free emission helps constrain the synchrotron spectrum at high frequencies.  Furthermore, the free-free emission has been proposed as an accurate star-formation rate indicator, since it is traces the ionizing photon generation rate \citep[e.g.,][]{Murphy11}.

I fit models to M82's radio spectrum, extracting information on these properties.  My primary purpose here is to demonstrate how low frequency radio spectra can be fit with the new free-free absorption models.  I ignore possible instrumental effects like beam sizes.  I also make no attempt to derive errors on the parameters, because my predicted fluxes have non-linear dependences on parameters.  Future studies can address these shortcomings.}

\subsection{Data for M82's Radio Spectrum}
The brightest starburst in the radio sky is M82.  Located $\sim 3.6\ \Mpc$ away (as adopted in this section from \citealt{Freedman94}, although \citealt{Sakai99} measure a distance of 3.9 Mpc), it has a total infrared luminosity of $5.9 \times 10^{10}\ \Lsun$ \citep{Sanders03}, corresponding to a Salpeter IMF star-formation rate of $10\ \Msun\ \yr^{-1}$ \citep{Kennicutt98}.  Most of the radio and infrared emission comes from a region of radius $\sim 250\ \pc$ \citep{Goetz90,Williams10}.  The starburst is viewed essentially edge-on from Earth.  Unlike the other bright starburst, NGC 253 \citep{Carilli96}, {\bfnop a large fraction} of the radio emission comes from the starburst itself rather than the host galaxy \citep{Klein88,Basu12,Adebahr12}.  This is especially important at low frequencies, where galaxies are frequently unresolved: the starburst may be obscured by its own H II regions, but the host galaxy is likely to be unobscured {\bfnop down to $\la 10\ \MHz$}.

{\bfnop I consider two sets of flux measurements.  One spans 22.5 MHz to 92 GHz where the starburst of M82 is unresolved.  The other includes the data of \citet{Adebahr12} where the starburst region itself is resolved and with frequencies 330 MHz to 10 GHz; I add high frequency flux measurements at 20 - 100 GHz to this data set.}

{\bfnop \emph{The total unresolved spectrum of M82} --} \citet{Williams10} obtained high quality radio observations in the {\bfnop 1 -- 7} GHz range using the Allen Telescope Array (ATA).  {\bfnop They find that the nonthermal radio spectrum is relatively flat ($S_{\nu} \propto \nu^{-0.6}$) and curved, becoming steeper at higher frequencies.}

They also compile interferometric and single-dish observations in the frequency range of 20 0- 100 GHz, which are useful in constraining free-free emission.  Noting {\bfnop an} offset {\bfnop towards greater flux} in the single-dish observations, \citet{Williams10} do not include single-dish observations in their modelling.  {\bfnop They argue that since the largest angular scale of the interferometric observations is $\sim 45\farcm$, the interferometric observations should not resolve out diffuse emission in the starburst core (with a diameter of $\sim 30\farcm$).  The greatest disparity between the interferometric and single-dish observations is at 23 GHz where the flux is $\sim 35\%$ greater in a Effelsberg 100 metre single dish observation than in a VLA observation.  I include the single-dish observations in my fits for two reasons: (1) the offset is fairly small compared to the uncertainty in the amount of free-free emission and (2) the high frequency data is relatively sparse.  Excluding the high frequency single-dish data does not change the total spectrum fit appreciably.}

\begin{table*}
\begin{minipage}{170mm}
\caption{Low Frequency Radio Data for M82's Total Spectrum}
\label{table:M82LowNuData}
\begin{tabular}{lcp{5cm}lp{5cm}}
\hline
$\nu$ & $S_{\nu}$       & Reference        & Synthesized beam size ($\theta$) $^a$      & Instrument\\
(MHz) & (Jy)            &                  &                                & \\
\hline
22.5 & $39 \pm 5$       & \citet*{Roger86}  & $66\farcm \times 102\farcm$    & Dipole array, Dominion Radio Astrophysical Observatory\\
38   & $23 \pm 3$       & \citet*{Kellermann69}$^b$  & $45\farcm \times 45\farcm\ {\rm sec} \zeta $ & Aperture-synthesis system, Mullard Radio Astronomy Observatory\\
57.5 & $29 \pm 6$       & \citet{Israel90} & $7\farcm0 \times 6\farcm5$     & Aperture-synthesis system, Clark Lake Radio Observatory\\
74   & $18.36 \pm 1.86$ & \citet{Cohen07}  & $80\farcs$ & VLA\\
86   & $27.8 \pm 3.8$   & \citet{Artyukh69,Laing80}  & $\sim 12\farcm \times 12\ \farcm$ & DKR-1000\\
151  & $16.82^c \pm 0.62$ & \citet{Baldwin85,Hales91} & $4\farcm2 \times 4\farcm2\ {\rm cosec}\,\delta$ & 6C aperture-synthesis telescope\\
178  & $15.3 \pm 0.7$   & \citet{Kellermann69}$^b$ & $23\farcm \times 18\farcm\ {\rm sec}\,\zeta$ & Aperture-synthesis system, Mullard Radio Astronomy Observatory\\
333  & $14 \pm 1$       & \citet{Basu12} & $22\farcs \times 15\farcs$       & GMRT\\
750  & $10.7 \pm 0.5$   & \citet{Kellermann69}$^b$ & $18\farcm5 \times 18\farcm5$    & Green Bank {\bfnop 300-foot transit} telescope\\
\hline
\end{tabular}
\\$^a$: The zenith angle is $\zeta$ and the declination is $\delta$.
\\$^b$: As recalibrated by \citet{Klein88}.
\\$^c$: Average of two measurements.\\
\end{minipage}
\end{table*}

Besides these observations, there are quite a few observations below 1 GHz, although the errors are naturally larger (see Table~\ref{table:M82LowNuData}).  In theory, the wide frequency coverage of the data, spanning from 22.5 MHz to 92 GHz, makes M82 a good choice for spectral modelling.  However, I note that many different instruments were used in collecting this data, with widely disparate beam sizes.  For most of the measurements below 1 GHz, not even the host galaxy (diameter $\sim 10\farcm$) is resolved.  In contrast, the ATA has a synthesized beam diameter of 4\farcm2 at 1 GHz and 35\farcs\ at 7 GHz, sufficient to resolve the host galaxy \citep{Williams10}.  The VLA and GMRT low frequency observations also had beam sizes small enough to resolve the host galaxy (and the starburst itself for the GMRT; \citealt{Cohen07,Basu12}).  The radio flux from the 74 MHz VLA sky survey is only $\sim 2/3$ of that from the unresolved 57.5 and 86 MHz measurements, which may mean that some flux is missing (perhaps from the host galaxy {\bfnop or radio halo}).  However, the error bars are very large, so it is unclear this is the case; furthermore, the 74 MHz VLA sky survey did report integrated fluxes even for resolved sources \citep{Cohen07}.  In any case, the 74 MHz VLA sky survey reports a major axis size for M82 of $66\farcs9 \pm 2\farcs8$ (smaller than the beam size), indicating that the majority of M82's 74 MHz radio emission comes from within 600 pc of its centre \citep{Cohen07}.  Thus, it appears the starburst itself is emitting at these frequencies, not just the host galaxy.  


For the total data set, I ignore the different beam sizes and assume all of the radio data points accurately measure the radio flux from the inner starbursting region of M82.  These results should therefore be treated cautiously.  I also {\bfnop redo} the fits {\bfnop to the total spectrum} using only the radio data points where the beam size was less than $10\farcm$, to see how measurements with large beam sizes were affecting my results.  

{\bfnop I note that \citet*{Marvil09} presented simple model fits to the unresolved frequency radio spectrum of M82 and other starbursts, including at low frequency.  They argued that the spectra indicated any free-free absorption must be inhomogeneous.}

{\bfnop \emph{Recent resolved observations} -- \citet{Adebahr12} recently presented resolved radio observations of M82 at frequencies $0.3 - 10\ \GHz$, seperating the starburst region emission from a `halo' component.  In contrast to \citet{Williams10}, \citet{Adebahr12} find an even flatter spectrum ($\nu^{-0.5}$) with no evidence of curvature.  The overall flux level is systematically lower than the \citet{Williams10} spectrum, even at 5 GHz where the \citet{Williams10} beam is small enough to resolve the starburst.  For this reason, I do a separate fitting of the \citet{Adebahr12} data points plus the high frequency ($> 10\ \GHz$) data points compiled in \citet{Williams10}.  For this spectrum, including the single dish observations is especially important since there are relatively few data points and many free parameters in my models.

In contrast to the unresolved low frequency emission, the 92 cm flux is significantly less than the 21 cm flux.  That implies strong free-free absorption.  Whether the absorption becomes stronger still at lower frequencies is impossible to tell at this time.

I focus on the \citet{Adebahr12} (and high frequency) data, because I am interested in the absorption properties of the starburst region itself.  I briefly discuss the fits to the total unresolved spectrum as well, though, for completeness.}

\subsection{Fitting Procedure}
I {\bfnop then} fit the properties of the H II regions {\bfnop and radio spectrum}.  {\bfnop A} star-formation rate sets the number of {\bfnop absorbing} H II regions and is allowed to be ${\rm SFR}_{\bfnop \rm eff} = 0.625, 1.25, 2.5, 5, 10, 20, {\bfnop 30}\ \Msun\ \yr$.  The electron density is allowed to be $n_e = 100, 200, 300, 600, 1200, 1800\ \cm^{-3}$, and the electron temperature can be $T = 5000, 7500, 10000, 12500, 15000, 20000\ \Kelv$.  These {\bfnop properties} set the Str\"omgren radius and turnover frequency for each H II region.  Finally, I set the scale height of the starburst $h_{\rm SB}$, which affects the covering fraction, to be {\bfnop 25, 50, or} 100 pc.  I consider all four scenarios described in Section~\ref{sec:Scenarios}.  For SSC models, I {\bfnop set the} low mass cutoff {\bfnop to} $1000\ \Msun$ and {\bfnop the} high mass cutoff {\bfnop to} $5 \times 10^6\ \Msun$.  In the aging SSC model, I assume M82's starburst is 15 Myr old \citep{Forster03}.  {\bfnop For all of these models, I compute $\Psi_{\rm SB}$, the fraction of {\bfnop emitted} flux transmitted by the starburst region after absorption by the H II regions, with equation~\ref{eqn:FRatioEdgeOnDisk} for an edge-on disc.  Appendix~\ref{sec:IntegratedFluxCalc} summarizes the calculation of $\Psi_{\rm SB}$ for different starburst geometries (edge-on and face-on discs and spheres).}  

The radio emission {\bfnop is} a combination of synchrotron and free-free emission.  Rather than running models of cosmic ray $e^{\pm}$ populations, a process {\bfnop that} would introduce many free parameters, I use the phenomenological function form of \citet{Williams10} for the unabsorbed synchrotron spectrum:
\begin{equation}
\log_{10} \left(\frac{S_{\rm nt}^{\rm unabs}}{\Jy}\right) =  {\cal A} + {\cal B} \log_{10} \left(\frac{\nu}{\GHz}\right) + {\cal C} \left[\log_{10} \left(\frac{\nu}{\GHz}\right)\right]^2
\end{equation}
In this formula, $-{\cal B}$ corresponds to the spectral index and ${\cal C}$ is the spectral curvature.

{\bfnop Any H II regions that contribute to free-free absorption must also emit free-free emission.  The free-free luminosity of a starburst is roughly
\begin{eqnarray}
\nonumber L_{\nu}^{\rm ff} & = & 2.2 \times 10^{25}\ \erg\ \sec^{-1}\ \Hz^{-1}\ \left(\frac{\rm SFR_{\rm eff}}{\Msun\ \yr^{-1}}\right)\\
& & \times \left(\frac{T}{10^4\ \Kelv}\right)^{0.45} \left(\frac{\nu}{\GHz}\right)^{-0.1}
\end{eqnarray}
using the $Q_{\rm ion}$ from equation~\ref{eqn:QwSFR}, at frequencies where the H II gas is transparent \citep{Condon92}.  The ${\rm SFR}_{\rm eff}$ is an `effective star-formation rate', which accounts for the escape of ionizing photons or their destruction by dust absorption.  But H II regions, being potentially very dense, can be individually optically thick at GHz frequencies, even not counting obscuration by other H II regions.  Therefore, I self-consistently calculate the luminosity of each H II region that contributes to absorption, using equation~\ref{eqn:HIIFFLuminosity}.  From these luminosity spectra, I can sum up the flux $S_{\rm min-ff}^{\rm unabs}$ that the collection of H II regions would have without absorption from the starburst.  I also calculate the covering fraction (equation~\ref{eqn:FCoverEdgeOnDisk}) and the filling factor (equation~\ref{eqn:fFill}).}

It is possible that there is ionized gas which contributes to the free-free emission but not to absorption.  This can happen if there is a population of small but dense H II regions with low covering fraction, such as ultracompact H II regions.  {\bfnop To represent this kind of contribution, I compute the free-free flux $S_{\rm dense-ff}$ from a population of H II regions of density $n_e = 10^4\ \cm^{-3}$ and temperature $T = 10^4\ \Kelv$.  I assume these H II regions surround `simple sources' with ionizing luminosities $Q_{\rm ion}^{\star} = 10^{49}\ \sec^{-1}$.  The amount of this emission is scaled by an effective star-formation rate (${\rm SFR}_{\rm add}$).  Since these H II regions are very small and do not occult much flux, I ignore their absorption effects on the radio spectrum (that is, they are not included in the calculation of $\Psi_{\rm SB}$).}

{\bfnop I also consider the effects of the host galaxy's WIM on the radio spectrum.  The WIM acts as a foreground screen, transmitting a fraction $\Psi_{\rm host} = e^{-\tau_{\rm ff}^{\rm host}}$ (where $\tau_{\rm ff}^{\rm host} = \alpha_{\rm H\ II} f_{\rm fill}^{\rm WIM} R_{\rm gal}$) of the starburst core flux through.  If M82's WIM is like the Milky Way's, it blocks essentially all emission from M82's centre at frequencies of a few MHz since M82 is viewed edge-on.  I treat the WIM as a medium with density $n_e = n_H = 0.2\ \cm^{-3}$ with a filling factor of $f_{\rm fill}^{\rm WIM} = 10\%$ and temperature of $10^4\ \Kelv$ \citep[c.f.,][]{Peterson02}.  The geometry is a disk with radius $R_{\rm gal} = 10\ \kpc$ and midplane-to-edge height of $1\ \kpc$.  I then find that the WIM becomes opaque ($\tau = 1$) at $\sim 5\ \MHz$.  In addition, there is free-free flux $S_{\rm WIM-ff}$ from the WIM, which is calculated self-consistently with equation~\ref{eqn:FRatioEdgeOnDisk}.  I include this component when considering the unresolved spectrum of M82, as it must be present at some level.  It is not included, however, when I fit to the resolved data, since only a small fraction of the host galaxy covers the starburst region itself.}

{\bfnop Finally, I calculate the absorption from the Milky Way WIM, specifically, the fraction of flux transmitted $\Psi_{\rm MW} = e^{-\tau_{\rm ff}^{\rm MW}}$ (where $\tau_{\rm ff}^{\rm host} = \alpha_{\rm H\ II} f_{\rm fill}^{\rm WIM} s$).  I assume the Galactic WIM has a density of $n_e = n_H = 0.225\ \cm^{-3}$, temperature of $7000\ \Kelv$, and a filling factor of $f_{\rm fill}^{\rm WIM} = 11\%$ \citep{Peterson02}.  The sightline along the WIM is $s = h_{\rm WIM} / \cos l$, where $h_{\rm WIM} = 0.83\ \kpc$ is the scale height of the WIM \citep{Peterson02} and $l = 40.567^{\circ}$ is M82's Galactic latitude.  I find that the Galactic WIM is opaque below 2.5 MHz; therefore M82's WIM is more important.}

The total predicted radio spectrum of M82 is then
\begin{equation}
S_{\nu}^{\rm pred} = {\bfnop [(S_{\rm nt}^{\rm unabs} + S_{\rm min-ff}^{\rm unabs} + S_{\rm dense-ff}^{\rm unabs}) \Psi_{\rm SB} \Psi_{\rm host} + S_{\rm WIM-ff}]\Psi_{\rm MW}}
\end{equation}
{\bfnop As I noted before, $S_{\rm WIM-ff}$ is only included when fitting the unresolved spectrum of M82.}

For each combination of parameters describing radio absorption (${\rm SFR}_{\rm eff}, n_e, T, h_{\rm SB}$), I use $\chi^2$ fitting to find the values of ${\cal A}$, ${\cal B}$, ${\cal C}$, ${\rm SFR}_{\rm add}$ that best fit the radio spectrum.  The values of ${\cal A}$ are allowed to range from 0 to 2, with a spacing of 0.01; I try values of ${\cal B}$ from -0.9 to -0.3, with a spacing of 0.02; ${\cal C}$ ranges from -0.2 to 0.1, with a spacing of 0.02; and $\log_{10} {\rm SFR}_{\rm add}$ ranges from -1 to 1 with a spacing of $\log_{10} 1.1$ plus the possibility of ${\rm SFR}_{\rm add} = 0$.  Then, I compare the $\chi^2$ values for the fits for each absorption parameter set to find the best-fitting model.

As a {\bfnop comparison} to these models, I also consider models where there {\bfnop is} no free-free absorption, and uniform slab models.  In each case, ${\cal A}$, ${\cal B}$, and ${\cal C}$ were free parameters.  For the no-absorption models, I simply add a free-free emission component parametrized by $ {\rm SFR}_{\rm add}$.  In the uniform slab models, the electron temperature $T$ {\bfnop is} a free parameter, with the same {\bfnop allowed} values as in the discrete H II region models.  I then {\bfnop choose} $n_e$ to fit the free-free emission and absorption.  In each case, I use $\chi^2$ fitting to the radio data to select the best-fitting parameters.

\subsection{Results for Unresolved Spectrum}

{\bfnop When I fit to the entire unresolved spectrum of M82, the discrete H II regions models} are better at reproducing the low frequency radio spectrum of M82 than either a fit without absorption or a uniform slab model (Table~\ref{table:M82Fits}).\footnote{\bfnop When fitting the uniform slab model to the unresolved spectrum, I considered only data with $\nu \ge 1\ \GHz$.  Including data with lower frequencies always resulted in a best-fitting model with $n_e = 0$.}   {\bfnop The flux is mostly transmitted in the models with discrete H II region, except at low frequency where the host galaxy's WIM acts as a screen.  Using these data, my conclusions are: (1) the unabsorbed synchrotron spectrum is fairly flat (${\cal B} \approx -0.6$) and negatively curved (${\cal C} \approx -0.1$); (2) the H II regions fill a very small fraction of the starburst volume ($\sim 0.3 - 3\%$) and cover only roughly $\sim 1/2$ of the region; (3) most ($\sim 60 - 75\%$) of the starburst's radio flux is transmitted at frequencies 10 - 100 MHz; (4) the pressures in the thermal H II regions are uncertain at an order of magnitude, but are quite high ($P_{\rm therm} / k \approx (0.2 - 2) \times 10^7\ \Kelv\ \cm^{-3}$); (5) the amount of free-free emission and absorption implies a low luminosity of ionizing photons, equivalent to a SFR of $\sim 1$ -- $2\ \Msun\ \yr^{-1}$.}

\begin{table*}
\begin{minipage}{170mm}
\caption{M82 Radio Spectrum Fits}
\label{table:M82Fits}
\begin{tabular}{lccccccccc}
\hline
Quantity & No         & Uniform & Simple  & \multicolumn{2}{c}{Stellar Population} & \multicolumn{2}{c}{Simple SSCs} & \multicolumn{2}{c}{Aging SSCs}\\
         & Absorption & Slab    & Sources &                                        &                                 &\\
\hline
\multicolumn{10}{c}{Unresolved spectrum}\\
Data     &  All$^a$      & $\ge \GHz$   & All$^a$          & All & $\theta < 10\farcm$ $^b$ & All & $\theta < 10\farcm$ $^b$  & All & $\theta < 10\farcm$ $^b$\\
\hline
${\rm SFR}_{\rm eff}$ ($\Msun\ \yr^{-1}$)$^c$          & \nodata & 0.61    & 0.625  & 0.625  & 0.625  & 0.625  & 0.625  & 0.625  & 1.25\\
$n_e$ ($\cm^{-3}$)                                     & \nodata & 25      & 600    & 300    & 600    & 200    & 300    & 200    & 100\\
$T_e$ ($\Kelv$)                                        & \nodata & 15000   & 20000  & 15000  & 20000  & 12500  & 20000  & 12500  & 7500\\
$h$ ($\pc$)                                            & \nodata & 100     & 100    & 100    & 100    & 50     & 100    & 25     & 100\\
${\cal A}$                                             & 0.94    & 0.95    & 0.94   & 0.95   & 0.94   & 0.96   & 0.94   & 0.96   & 0.95\\
${\cal B}$                                             & -0.56   & -0.58   & -0.58  & -0.60  & -0.58  & -0.62  & -0.58  & -0.62  & -0.60\\
${\cal C}$                                             & -0.20   & -0.06   & -0.12  & -0.08  & -0.14  & -0.08  & -0.14  & -0.10  & -0.10\\
${\rm SFR}_{\rm add}$ ($\Msun\ \yr^{-1}$)$^e$          & 3.40    & \nodata & 1.44   & 0.98   & 1.91   & 1.19   & 1.91   & 2.11   & 1.74\\
$\chi^2$                                               & 114     & 72.9$^k$& 87.4 (79.0) & 83.5   & 77.4   & 86.3   & 78.6   & 84.1   & 77.5\\
$f_{\rm cover}$                                        & \nodata & 100\%   & 43\%   & 60\%   & 36\%   & 52\%   & 25\%   & 59\%   & 57\%\\
$f_{\rm fill}$                                         & \nodata & 100\%   & 0.27\% & 0.69\% & 0.22\% & 1.7\%  & 0.57\% & 2.1\%  & 2.8\%\\
$f_{\rm therm}^{\rm SB}$ (1 GHz)$^f$                   & 1.3\%   & 1.9\%   & 2.8\%  & 2.2\%  & 2.8\%  & 2.0\%  & 2.8\%  & 1.4\%  & 1.8\%\\
$\Psi_{\rm SB} (\nu \to 0)$ $^g$                       & \nodata & 0\%     & 74.6\% & 62.5\% & 79.4\% & 60.7\% & 81.7\% & 63.6\% & 64.9\%\\
$\Psi_{\rm SB} (1\ \GHz)$ $^h$                         & \nodata & 97.4\%  & 97.8\% & 97.5\% & 97.9\% & 94.8\% & 97.9\% & 95.9\% & 97.3\%\\
$(\alpha_{\rm eff})^{-1} (\nu \to 0)$ (pc) $^i$        & $\infty$ & 0      & 680    & 410    & 880    & 380    & 1000   & 420    & 450\\
$P_{\rm therm} / k$ ($\Kelv\ \cm^{-3}$) $^j$           & \nodata & $7.5 \times 10^5$ & $2.4 \times 10^7$ & $9.0 \times 10^6$ & $2.4 \times 10^7$ & $5.0 \times 10^6$ & $1.2 \times 10^7$ & $5.0 \times 10^6$ & $1.5 \times 10^6$\\
\hline
\multicolumn{10}{c}{Core spectrum}\\
Data     &  All      & All   & All          & \multicolumn{2}{c}{All}      & \multicolumn{2}{c}{All}  & \multicolumn{2}{c}{All}\\
\hline
${\rm SFR}_{\rm eff}$ ($\Msun\ \yr^{-1}$)$^c$          & \nodata & 0.48    & 0.625  & \mcc{0.625}  & \mcc{0.625}  & \mcc{2.5}\\
$n_e$ ($\cm^{-3}$)                                     & \nodata & 45      & 300    & \mcc{200}    & \mcc{100}    & \mcc{200}\\
$T_e$ ($\Kelv$)                                        & \nodata & 15000   & 12500  & \mcc{12500}  & \mcc{12500}  & \mcc{15000}\\
$h$ ($\pc$)                                            & \nodata & 25      & 25     & \mcc{25}     & \mcc{25}     & \mcc{25}\\
${\cal A}$                                             & 0.71    & 0.83    & 0.84   & \mcc{0.84}   & \mcc{0.83}   & \mcc{0.84}\\
${\cal B}$                                             & -0.32   & -0.54   & -0.56  & \mcc{-0.56}  & \mcc{-0.54}  & \mcc{-0.58}\\
${\cal C}$                                             & -0.20   & -0.02   & -0.02  & \mcc{-0.02}  & \mcc{-0.04}  & \mcc{-0.02}\\
${\rm SFR}_{\rm add}$ ($\Msun\ \yr^{-1}$)$^e$          & 2.81    & \nodata & 0.16   & \mcc{0.19}   & \mcc{0.46}   & \mcc{0.10}\\
$\chi^2$                                               & 31.3    & 9.17    & 9.25   & \mcc{9.26}   & \mcc{9.26}   & \mcc{9.49}\\
$f_{\rm cover}$                                        & \nodata & 100\%   & 96.6\% & \mcc{98.2\%} & \mcc{95.2\%} & \mcc{95.8\%}\\
$f_{\rm fill}$                                         & \nodata & 100\%   & 2.4\%  & \mcc{5.2\%}  & \mcc{13.0\%} & \mcc{9.5\%}\\
$f_{\rm therm}^{\rm SB}$ (1 GHz)$^f$                   & 1.8\%   & 2.0\%   & 3.1\%  & \mcc{3.1\%}  & \mcc{3.3\%}  & \mcc{4.8\%}\\
$\Psi_{\rm SB} (\nu \to 0)$ $^g$                       & \nodata & 0\%     & 21.5\% & \mcc{16.4\%} & \mcc{18.7\%} & \mcc{23.4\%}\\
$\Psi_{\rm SB} (1\ \GHz)$ $^h$                         & \nodata & 92.0\%  & 89.2\% & \mcc{89.5\%} & \mcc{89.3\%} & \mcc{85.9\%}\\
$(\alpha_{\rm eff})^{-1} (\nu \to 0)$ (pc) $^i$        & $\infty$ & 0      & 88     & \mcc{66}     & \mcc{75}     & \mcc{96}\\
$P_{\rm therm} / k$ ($\Kelv\ \cm^{-3}$) $^j$           & \nodata & $1.3 \times 10^6$ & $7.5 \times 10^6$ & \mcc{$5.0 \times 10^6$} & \mcc{$2.5 \times 10^6$} & \mcc{$6.0 \times 10^6$}\\
\hline
\end{tabular}
\\$^a$: The best-fitting parameters for these models are the same whether all data are included or just those with beam sizes less than 10 arcminutes.  The $\chi^2$ value when only data with beam sizes less than 10 arcminutes is given in parentheses.
\\$^b$: Fluxes for observations with beam sizes less than 10 arcminutes (see Table~\ref{table:M82LowNuData}).  
\\$^c$: Effective star-formation rate, which sets the ionizing photon luminosity of the starburst.  This sets both the number of H II regions and the amount of free-free emission.
\\$^d$: Thermal free-free emission at 1 GHz from the H II regions responsible for free-free absorption.  This sets a floor on the amount of free-free emission.
\\$^e$: Additional free-free emission from ionized gas that does not contribute to free-free absorption in my model (e.g., from more compact, denser H II regions).  This is a free parameter.
\\$^f$: Fraction of the flux {\bfnop from the starburst} at 1 GHz {\bfnop that} is thermal free-free emission.
\\$^g$: Fraction of the radio flux transmitted by the free-free absorbing H II regions at extremely low frequencies.  The effects of free-free absorption in other phases or synchrotron self-absorption are not included.
\\$^h$: Fraction of the radio flux transmitted by the free-free absorbing H II regions at 1 GHz.
\\$^i$: {\bfnop Typical length on a sightline through the starburst before hitting an H II region.}
\\$^j$: Derived thermal pressure in H II regions, ${\bfnop 2} n_e T_e$.
\\$^k$: In selecting the uniform slab model, $\chi^2$ was minimized for data above 1 GHz; this value is {\bfnop 72.9}.  However, if {\bfnop the entire unresolved radio spectrum} is included, the total $\chi^2$ increases to {\bfnop 578}.
\end{minipage}
\end{table*}

When I fit only the data with beam sizes smaller than 10\farcm, the basic conclusion that most of the flux is transmitted stands.  The best-fitting {\bfnop simple source} model is {\bfnop the same for} these data.  However, the models of Str\"omgren spheres {\bfnop around a stellar population} and around SSCs now fit better for different {\bfnop densities and temperatures}.  Therefore, the environments of the H II regions are not well constrained.

\subsection{Results for \citet{Adebahr12} Spectrum}
\label{sec:A12Results}
\begin{figure}
\centerline{\includegraphics[width=9cm]{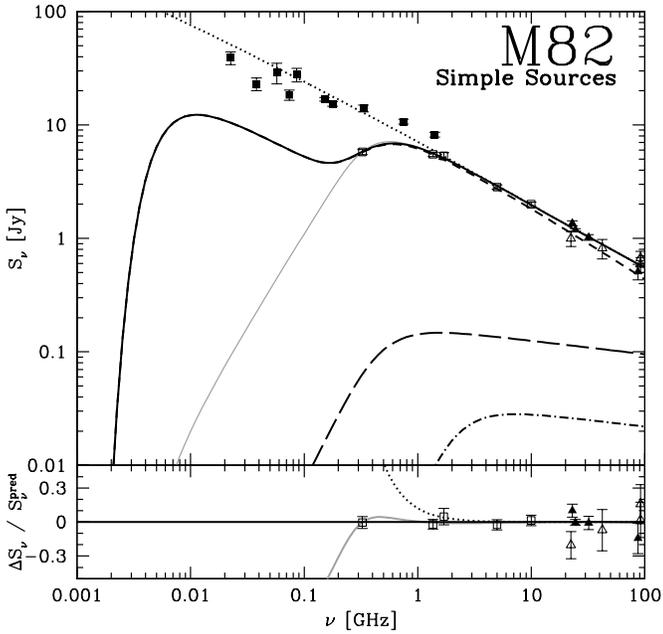}}
\caption{The effects of free-free absorption on M82's radio spectrum.  This is the best-fitting model to the \citet{Adebahr12} and high-frequency data for H II regions surrounding simple sources.  The total absorbed spectrum is shown as the black solid line.  Its components are the synchrotron emission (short-dashed), free-free emission from absorbing H II regions (long-dashed), and free-free emission from dense non-absorbing H II regions (dash-dotted).  The dotted line is the total emission if there were no absorption.  I show a uniform slab fit to the $\ge 1\ \GHz$ radio spectrum as the grey solid line.  The residuals for the resolved data points are plotted on the bottom. The open squares are the \citet{Adebahr12} data and the triangles are the high frequency data compiled in \citet{Williams10} (open for interferometric and filled for single-dish), while the filled squares are the low frequency unresolved observations in Table~\ref{table:M82LowNuData}. \label{fig:FitM82RadioSpectrumIndiv}}
\end{figure}

\begin{figure}
\centerline{\includegraphics[width=9cm]{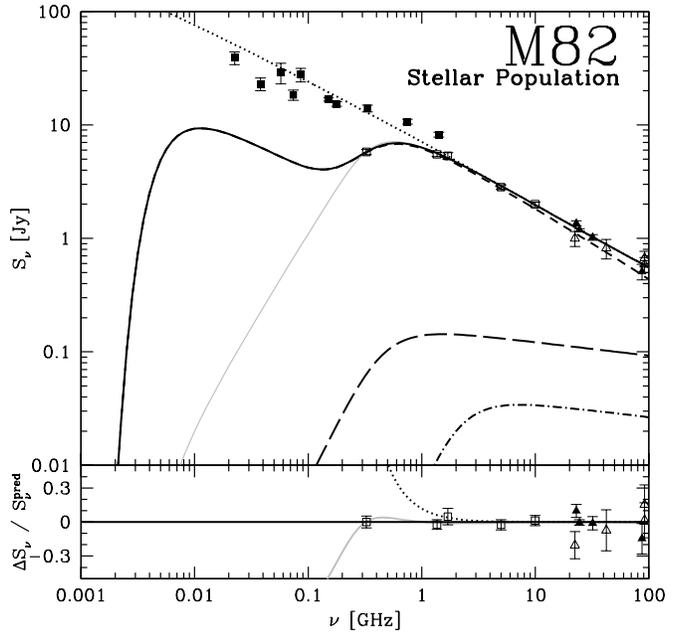}}
\caption{The effects of free-free absorption on M82's radio spectrum, for H II regions surrounding a realistic stellar population.  The line and point styles are the same as in Figure~\ref{fig:FitM82RadioSpectrumIndiv}. \label{fig:FitM82RadioSpectrumPI}}
\end{figure}

\begin{figure}
\centerline{\includegraphics[width=9cm]{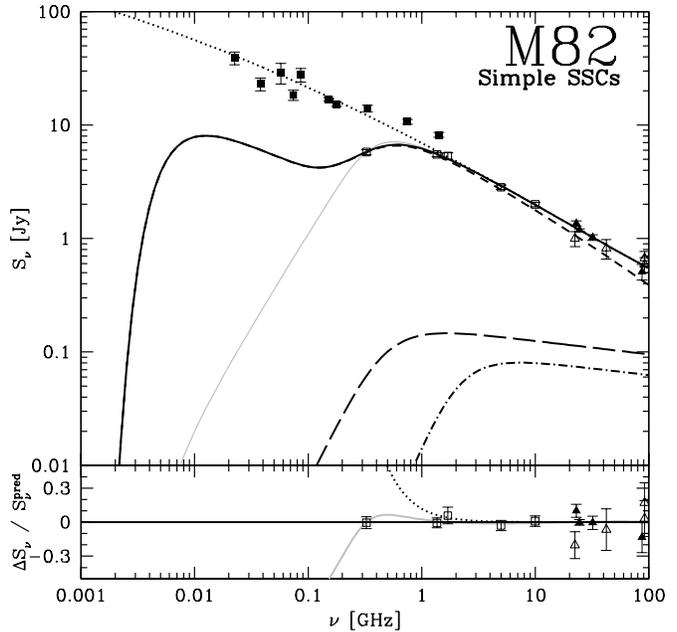}}
\caption{A model of the free-free absorption on M82's radio spectrum, using the simple super star cluster assumptions described in section~\ref{sec:SSCs}.  The line and point styles are the same as in Figure~\ref{fig:FitM82RadioSpectrumIndiv}. \label{fig:FitM82RadioSpectrumClust}}
\end{figure}

\begin{figure}
\centerline{\includegraphics[width=9cm]{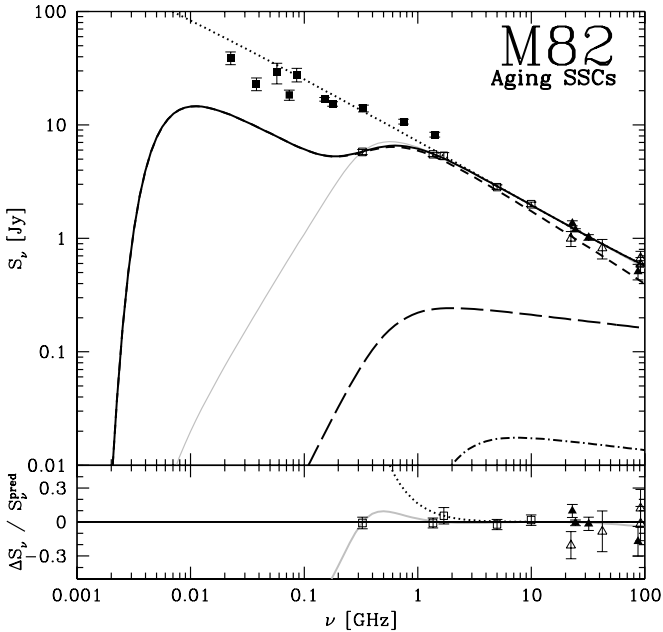}}
\caption{A model of the free-free absorption on M82's radio spectrum, {\bfnop accounting} for how the ionizing photon luminosity of SSCs evolve as they age.  The line and point styles are the same as in Figure~\ref{fig:FitM82RadioSpectrumIndiv}. \label{fig:FitM82RadioSpectrumAC}}
\end{figure}

{\bfnop My results are very different for the \citet{Adebahr12} spectrum.  I show the best-fitting models in Figure~\ref{fig:FitM82RadioSpectrumIndiv} (for simple sources), Figure~\ref{fig:FitM82RadioSpectrumPI} (for individual stars in a realistic stellar population), Figure~\ref{fig:FitM82RadioSpectrumClust} (for simple SSCs), and Figure~\ref{fig:FitM82RadioSpectrumAC} (for aging SSCs).  Overall, the best-fitting properties of the H II regions are similar in all four cases, except that the density is higher when the Str\"omgren spheres surround individual stars.  

{\bfnop \emph{Basic properties of the best-fitting models --}} Phenomenologically speaking, there are several conclusions that are similar in each of the four cases:}

\begin{itemize}
\item The unabsorbed synchrotron spectrum is fairly flat, with an intrinsic nonthermal spectral index of $\sim 0.55$, which is lower than the typical values of $\sim 0.8 - 0.9$ for normal spiral galaxies \citep[e.g.,][]{Niklas97}.  {\bfnop Even the uniform slab model has a similar spectral index; compare with the value of $\sim 0.5$ derived by \citet{Adebahr12}.}  My values are also consistent with the fit of \citet{Williams10}, ${\cal B} = -0.56 \pm 0.02$.  

\item The unabsorbed synchrotron spectrum {\bfnop shows little evidence of curvature}.  The derived intrinsic nonthermal spectral curvature in the best-fitting models is consistently {\bfnop ${\cal C} \approx -0.02$ to $-0.04$, including in the uniform slab model}.  {\bfnop The best-fitting curvature is substantially less than that \bfnop found by} \citet{Williams10}, ${\cal C} = -0.12 \pm 0.03$, {\bfnop but agrees with the results of \citet{Adebahr12}.}

\item The covering fraction of the H II regions is {\bfnop nearly unity}, and the filling fraction ranges from {\bfnop $2$ to $13\%$}.  As a result, most of the radio flux is {\bfnop absorbed at low frequencies.  Yet, roughly $\sim 15$ -- $20\%$ is transmitted even at frequencies of 10 to 100 MHz.  There is roughly $(\alpha_{\rm eff})^{-1} \sim 70$ -- $100\ \pc$ of synchrotron-emitting material on a typical sightline before intercepting an H II region.}  At 1 GHz, free-free absorption reduces the observed flux by {\bfnop about $10\%$}.  

{\bfnop \item Low electron densities in the H II regions and small scale heights are preferred by the fits.  Both of these trends lead to bigger covering fractions and stronger free-free absorption at low frequencies.}

\item These models {\bfnop -- including the uniform slab and no starburst absorption models --} require low ionizing photon luminosities relative to M82's star-formation rate of $10\ \Msun\ \yr^{-1}$.  {\bfnop This seems necessary to fit the falling (and therefore synchrotron dominated) spectrum at frequencies approaching 100 GHz.}  {\bfnop In most models,} the ionizing photon luminosity powering the absorbing H II regions is equivalent to a star-formation rate of only $\la 1\ \Msun\ \yr^{-1}$, {\bfnop despite including the free additional emission component from dense H II regions.  The ionizing photon luminosity is at most $3\ \Msun\ \yr^{-1}$ in the aging SSC and unabsorbed models.}  The thermal fraction at 1 GHz in these models is only {\bfnop $\sim$ 2 -- 5\%}.  
\end{itemize}

In contrast to the discrete H II region models, the best-fitting model with no free-free absorption (solid grey line in Figures~\ref{fig:FitM82RadioSpectrumIndiv} - \ref{fig:FitM82RadioSpectrumAC}) requires strong intrinsic spectral curvature ({\bfnop ${\cal C} = -0.20$}) and a flatter synchrotron spectrum ({\bfnop ${\cal B} = -0.32$}) in order to not overproduce the observed low frequency emission.  

{\bfnop Unlike the data in the total unresolved spectrum, the 333 MHz data point of \citet{Adebahr12} is fully consistent with the uniform slab model.  It has a slightly lower $\chi^2$, using fewer parameters (since there is no additional free-free component and the effective SFR is directly related to the density.  For practical purposes, the predicted spectra are essentially identical at frequencies above 333 MHz, and all the models are within the errors of the data points (compare the grey and black lines in Figures~\ref{fig:FitM82RadioSpectrumIndiv} -- \ref{fig:FitM82RadioSpectrumAC}).  Where the uniform slab and discrete H II region models diverge is below 300 MHz, as the uniform slab model continues to plummet while the discrete H II region model flattens again.  However, the no absorption model is a very poor fit.}

\begin{figure*}
\centerline{\includegraphics[width=9cm]{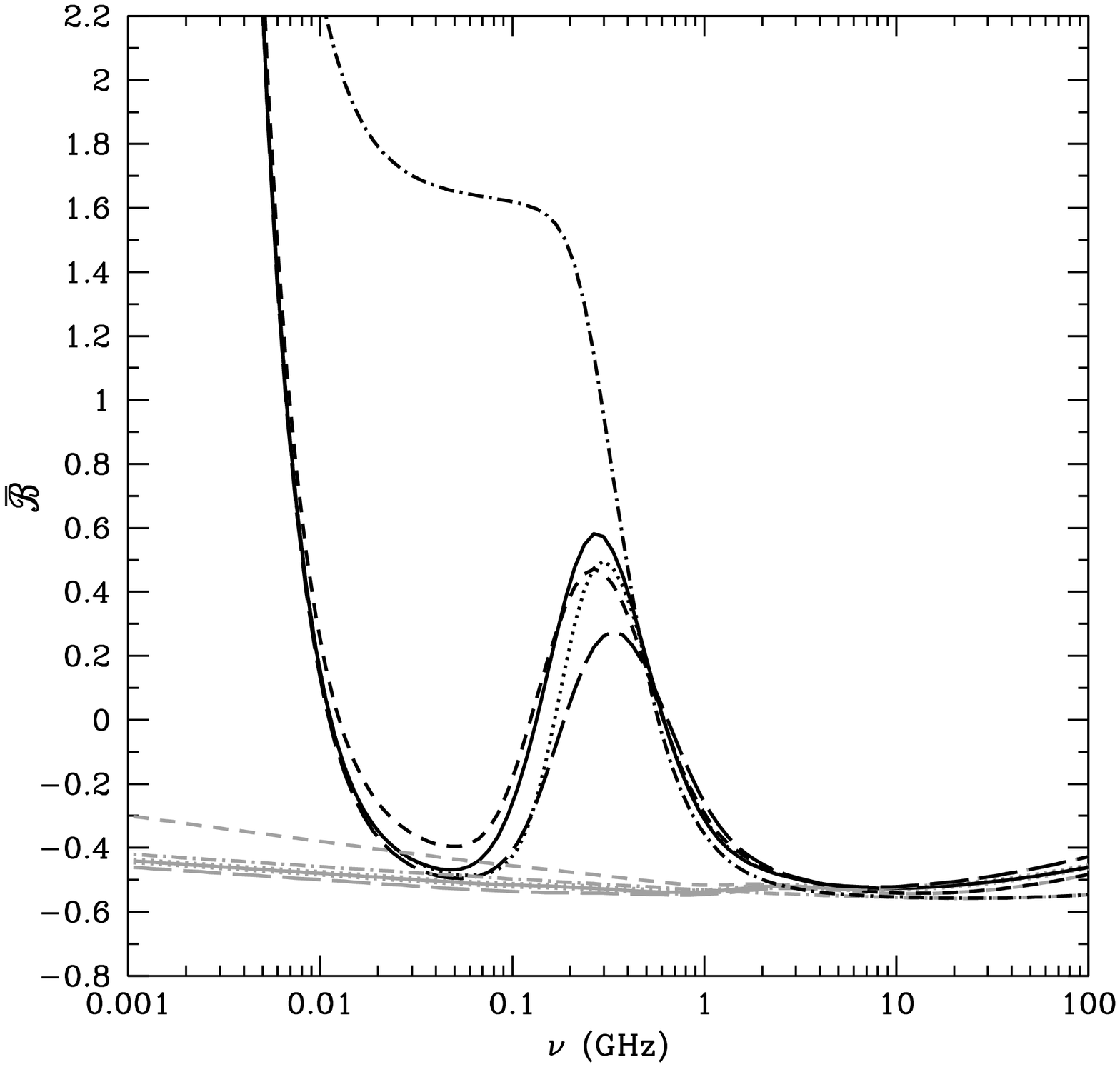}\includegraphics[width=9cm]{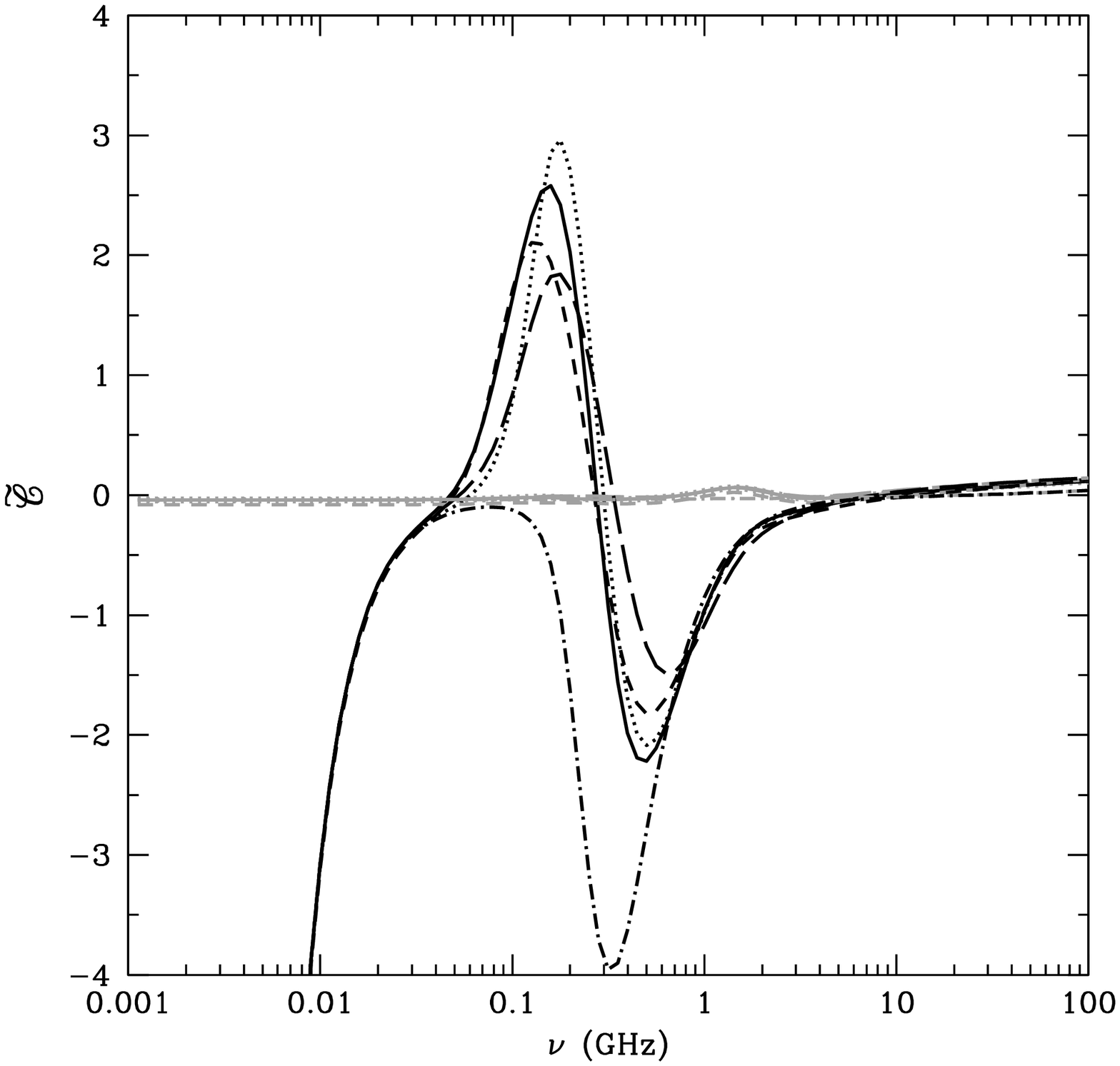}}
\caption{How free-free absorption affects the total spectral index $\tilde{\cal B}$ (left) and total spectral curvature $\tilde{\cal C}$ (right).  I show these values with (black) and without (grey) free-free absorption.  The line styles are for different {\bfnop best-fitting} M82 models; dotted is {\bfnop simple sources, solid is for a stellar population, short-dashed is simple SSCs, long-dashed is aging SSCs, and dash-dotted is the uniform slab model}.  Free-free absorption introduces a pulse in the spectral index and curvature, as H II regions transition from being transparent to opaque with decreasing frequency.\label{fig:FitM82SpectIndex}}
\end{figure*}

{\bfnop \emph{Effects on the spectrum shape --}} The effects of free-free absorption {\bfnop from discrete H II regions} on the spectrum shape are concentrated within a finite frequency band.  At high frequencies, the H II regions are transparent.  At low frequencies, the H II regions are completely opaque, and $\alpha_{\rm eff}$ reaches some constant value; this changes the normalization of the spectrum, but not its intrinsic shape.  These effects can be described more quantitatively in terms of the total spectral index $\tilde{\cal B}$ and spectral curvature $\tilde{\cal C}$:
\begin{eqnarray}
\tilde{\cal B} & \equiv & \frac{d \log_{10} S_{\nu}^{\rm pred}}{d \log_{10} \nu}\\
\tilde{\cal C} & \equiv & \frac{d^2 \log_{10} S_{\nu}^{\rm pred}}{d (\log_{10} \nu)^2};
\end{eqnarray}
these quantities differ from ${\cal B}$ and ${\cal C}$ in that they include free-free emission and absorption.\footnote{Note the use of base 10 logarithms.  While $\tilde{\cal B} = d\ln S_{\nu}^{\rm pred} / d\ln \nu$, $\tilde{\cal C} = \ln 10 \times d^2\ln S_{\nu}^{\rm pred} / d(\ln \nu)^2$}  I show how $\tilde{\cal B}$ and $\tilde{\cal C}$ vary with frequency in Figure~\ref{fig:FitM82SpectIndex} for the best-fitting models.  Free-free absorption (the difference between black and grey lines) flattens the spectrum (higher $\tilde{\cal B}$) between $\sim 0.1 - 1\ \GHz$: this is the regime where H II regions become optically thick.  The resulting pulse in $\tilde{\cal B}$ is narrower when H II regions surround simple sources (dotted lines) instead of SSCs ({\bfnop dashed lines}) or a realistic stellar population ({\bfnop solid line}): in that model, the H II regions all have the same size, with the same turnover frequency, instead of the range of turnover frequencies for SSCs {\bfnop or a stellar population.  In contrast, $\tilde{\cal B}$ plateaus at a value $\sim 1.6$ in the uniform slab model (dash-dotted line), as the free-free optical depth continues to rise.  Finally, at frequencies below 20 MHz, the host galaxy's WIM starts acting as a foreground screen to the starburst, and in all models $\tilde{\cal B}$ rises dramatically.}

{\bfnop Likewise, there is a pulse in $\tilde{\cal C}$ that flips between negative (at low frequencies) and positive (at high frequencies) in the discrete H II absorption model.  Free-free absorption in the uniform slab model, by contrast, causes a purely negative curvature.}  One {\bfnop obstacle} for studies {\bfnop that measure} intrinsic nonthermal curvature in the radio spectrum is that either free-free absorption or emission {\bfnop affects the spectrum at} most observable radio frequencies.  Only at frequencies of a few GHz does the intrinsic curvature dominate over that introduced by free-free absorption or emission.  {\bfnop Even at low frequencies, where $\alpha_{\rm eff}$ is constant, the host galaxy WIM starts forcing the curvature to negative values.}

{\bfnop \emph{The thermal pressure of H II regions --}} In my best-fitting {\bfnop discrete H II region} models, the {\bfnop thermal} pressure varies by a factor $\sim 3$, {\bfnop spanning the range $P_{\bfnop \rm therm} / k = (2.5 - 7.5) \times 10^6\ \Kelv\ \cm^{-3}$.}  Numerous studies have inferred the pressure of M82's H II regions, using infrared and optical spectroscopy.  \citet{Smith06} found a fairly high pressure of $P_{\bfnop \rm therm}/k = (1 - 2) \times 10^7\ \Kelv\ \cm^{-3}$ for the H II region around the M82 A-1 super star cluster, {\bfnop several times greater than in my models}.  Note that M82 A-1 is relatively large ($\sim 10^6\ \Msun$) compared to the small SSCs that would be expected to dominate the free-free absorption.  \citet{Westmoquette07} found lower pressures of $P_{\bfnop \rm therm}/k = (5 - 10) \times 10^6\ \Kelv\ \cm^{-3}$ in most M82 H II regions, {\bfnop comparable to those in my best-fitting models}.  \citet{Lord96} inferred H II region pressures of $P_{\bfnop \rm therm}/k = 3 \times 10^6\ \Kelv\ \cm^{-3}$, based on infrared diagnostics of the surrounding denser and colder photodissociation regions.  In any case, though, because of the range of densities I allowed for discrete H II regions, the pressures are necessarily higher than what I would find with a uniform slab model ($\sim 10^6\ \Kelv\ \cm^{-3}$; compare with the similar results for NGC 253 in \citealt{Carilli96}).  

{\bfnop \emph{Is free-free emission a good SFR indicator? --}} The low ionizing photon luminosities I derive are worrying, especially since I allow for `hidden' free-free emission from compact H II regions that do not contribute to absorption.  {\bfnop Even given} the large errors and sparseness of the high frequency data, {\bfnop it seems difficult for there to be large amounts of free-free emission}.  On the other hand, thermal dust emission probably contributes to the highest frequency data points at some level, and there may be spinning dust emission as well, but including any dust emission would just tighten the constraints on the free-free emission.  

It is possible, though, that the amount of free-free emission really is low in starburst galaxies -- perhaps not a surprising hypothesis given that starbursts are dusty places {\bfnop that could readily absorb ionizing photons}  \citep*[c.f.,][]{Petrosian72}.  Or, {\bfnop perhaps the problem is the opposite:} ultraviolet photons {\bfnop could easily} escape {\bfnop through} the hot superwind phase that likely fills much of the starburst volume {\bfnop \citep[e.g.,][]{Heckman90}}, instead of ionizing the molecular gas.  If so, then free-free emission is not necessarily a simple star-formation rate indicator in starbursts: its strength depends somehow on the radiative transfer {\bfnop in their} dusty, multiphase environments, and it is buried even more than expected by synchrotron emission. 

\subsection{Useful Future Data}

{\bfnop The need for high-resolution low frequency data is obvious.  With current data, it is impossible to meaningfully favor the uniform slab model or the discrete H II region model on observational grounds.   Aside from the single data point in \citet{Adebahr12} and the maps of \citet{Basu12}, the low frequency data so vital in constraining free-free absorption comes from many instruments with varying but poor
 beam sizes.  A future low frequency survey of M82 that resolves the starburst would ensure that flux from the surrounding galaxy is not making M82's starburst appear brighter than it really is; such a survey can be done with the GMRT, VLA, or LOFAR. }

Besides better low frequency data, it would be helpful to have high quality observations at 10 - 100 GHz.  Then {\bfnop the flattening from free-free emission can be better constrained}, if it is present.  The Jansky VLA is able to observe at up to 50 GHz \citep{Perley11}, and with its high sensitivity, {\bfnop it can measure the starburst spectrum with high accuracy}.  Moreover, as a single instrument, there would be fewer worries about systematics when comparing between data at different frequencies.  Furthermore, it has high spatial resolution, so it can actually measure the characteristic sizes of H II regions, in particular searching for small, high density H II regions.  

Another {\bfnop set of useful diagnostics comes from} radio recombination lines, also observable with the Jansky VLA \citep{Kepley11}.  {\bfnop These} constrain not only the ionizing photon luminosity of a starburst, but {\bfnop the densities and temperatures of the ionized gas}.  I note that \cite{Rodriguez-Rico04} found{\bfnop,} using H53$\alpha$ and H92$\alpha$ radio recombination lines{\bfnop,} an ionizing photon luminosity of $1.6 \times 10^{53}\ \sec^{-1}$ for M82{\bfnop.   That is} equivalent to a star-formation rate of $0.7\ \Msun\ \yr^{-1}$, compatible with the low ionizing photons I find here.  On the other hand, \citet{Puxley89} deduced an ionizing luminosity of $1.1 \times 10^{54}\ \sec^{-1}$ also using the H53$\alpha$ line, which would correspond to a star-formation rate of $5\ \Msun\ \yr^{-1}$.  This is significantly greater than what I find with my fits, but still a factor of $\sim 2$ lower than the infrared-derived star-formation rate.  Future radio recombination line studies may clarify the situation.

\section{Where do the radio spectra of starbursts really end?}
\label{sec:StarburstRadioEnd}
{\bfnop My models of the radio spectrum are valid if the only relevant absorption processes are free-free absorption from the starburst H II regions, the host galaxy WIM, and the Galactic WIM.  But the} starburst radio spectrum cannot continue indefinitely to zero frequency, {\bfnop even ignoring the effects of the host galaxy and Milky Way WIM}.  The presence of volume-filling phases of ionized gas within starbursts {\bfnop ensures} free-free absorption, {\bfnop as well as the Razin suppression of synchrotron emission}.  In addition, the purely nonthermal process of synchrotron self-absorption eventually must cut off the radio spectrum of starburst galaxies even if there is no diffuse ionized gas.  

{\bfnop Aside from limiting the applicability of my models, these effects might inform us on different aspects of starburst environments.  Free-free absorption from volume-filling ionized gas would probe the physical conditions in most of the starburst ISM.  Detection of the synchrotron self-absorption or Razin effect would constrain starburst magnetic fields.  They would also inform us of which ISM phase the synchrotron-emitting CR $e^{\pm}$ are in.  In this section, I check whether these processes have any effects on starburst radio spectra at relevant frequencies.}

\subsection{Is there a WIM in starburst regions?}
\label{sec:StarburstWIM}
{\bfnop 
The conclusion that starbursts are partly transparent at low frequencies requires that there be no volume-filling warm ionized medium within the starburst region.  Yet a WIM exists in the Milky Way and other normal star-forming galaxies, pervading at least 10\% of the Galaxy out to a scale height of 1 kpc (the properties and theory of the WIM in normal galaxies is reviewed in \citealt{Haffner09}).  The Milky Way WIM becomes opaque at frequencies of a few MHz \citep{Hoyle63,Alexander69}.  While it is not fully certain how ionizing photons traverse such large distances in Milky Way, current theory is that not all ionizing photons are confined to dense H II regions but propagate into underdensities in the inhomogeneous ISM (e.g., \citealt*{Ciardi02}; \citealt{Wood10}).   Might a volume-filling WIM form in starburst regions in a similar way and account for the observed radio absorption?

There are several reasons to expect this is not the case, at least not within the starburst region proper.  Firstly, other phases are expected to fill the starburst ISM, leaving little room for a WIM.  A low density, hot plasma fill starbursts where supernova remnants do not experience strong radiative losses and the pressure is not extreme, simply because the supernova rate density is so great \citep[e.g.,][]{Chevalier85,Heckman90,Lord96,Strickland07,Strickland09}.  \emph{Chandra} has detected iron line emission consistent with the existence of this plasma in M82 \citep{Strickland07,Strickland09}.  While ionized, this phase is so hot ($\sim 10^8\ \Kelv$) that there is essentially no free-free absorption (see section~\ref{sec:StarburstHotWind}).  Ionizing photons that escape into this phase propagate freely, but if the phase is volume-filling, they could easily escape the starburst entirely without ionizing neutral gas.  

In the more extreme starbursts like those within Arp 220, the molecular medium with most of the gas mass is so dense that supernova remnants probably stall and the hot phase cannot form (e.g., \citealt{Thornton98}; \citealt{Thompson05}).  Cold molecular gas likely fills most of these starburst regions.  While the large neutral mass would become a formidable WIM if it were ionized, its sheer density poses a nearly insurmountable hurdle for a volume-filling homogeneous WIM.  The neutral gas columns of starbursts are larger than those of the Milky Way by orders of magnitude.  This means that the average optical depth of the neutral gas is hundreds, or thousands, of times larger than the Milky Way, with neutral gas and dust drastically limiting the range of ionizing photons.  

The rapid recombination rate in the dense ISM is another obstacle for WIM formation.  The maximum volume that can be ionized is equivalent to a Str\"omgren volume $Q_{\rm ion} / (n_H \alpha_B)$.  The maximum fraction of the starburst that can be ionized is:
\begin{equation}
\label{eqn:fWIM}
f_{\rm fill} \le 0.029 \left(\frac{n_H}{1000\ \cm^{-3}}\right)^{-2} \left(\frac{\rho_{\rm SFR}}{1000\ \Msun\ \yr^{-1}\ \kpc^{-3}}\right).
\end{equation}
at $T = 10^4\ \Kelv$.  I list the values for the specific examples of the Galactic Centre Central Molecular Zone (CMZ), M82, and Arp 220 in Table~\ref{table:StarburstWIMConstraints}.  In each case, only a few percent of the molecular gas can be ionized, suggesting the WIM is confined to small bubbles within the cold molecular medium.

Could we get around this limit by supposing that a volume-filled WIM has low density?  The main problem is the surrounding ISM pressure, which is several orders of magnitude higher than in the Milky Way.  If the WIM is supported by thermal pressure, its minimum hydrogen atom density at the known pressure of a starburst region is $n_{\rm min} \ge P_{\rm ISM} / (2 k T)$.  The large results are shown in Table~\ref{table:StarburstWIMConstraints}.  Then through equation~\ref{eqn:fWIM}, thermally-supported WIM makes up a small fraction of the starburst volume.

\begin{table*}
\begin{minipage}{170mm}
\caption{Starburst WIM Constraints}
\label{table:StarburstWIMConstraints}
\begin{tabular}{lcccc}
\hline
Property                                          & Galactic Centre CMZ & M82 & Arp 220 Nuclei & References\\
\hline
${\rm SFR}$ ($\Msun\ \yr^{-1}$)                   & 0.07                & 10              & 100             & (1)\\
$R_{\rm SB}$ ($\pc$)                              & 112                 & 250             & 100             & (2)\\
$P_{\rm ISM} / k$ ($\Kelv\ \cm^{-3}$)             & $1.5 \times 10^6$   & $2 \times 10^7$ & $8 \times 10^9$ & (3)\\
$\Sigma_{\rm SFR}$ ($\Msun\ \yr^{-1}\ \kpc^{-2}$) & 2                   & 50              & 3000            & \nodata\\
$\rho_{\rm SFR}$ ($\Msun\ \yr^{-1}\ \kpc^{-3}$)   & 21                  & 510             & 32000           & (4)\\
\hline
\multicolumn{5}{c}{WIM carved from mean density gas}\\
\hline
$\mean{n_H}$ ($\cm^{-3}$)                         & 120                 & 410             & 10000        & (5)\\
$f_{\rm fill}$                                    & $\le 4.2\%$         & $\le 8.7\%$     & $\le 0.92\%$ & \nodata\\
\hline
\multicolumn{5}{c}{Thermally supported WIM}\\
\hline
$n_{\rm min}$ ($\cm^{-3}$) $^a$     & $75$         & $1000$       & $400000$                    & \nodata\\
$\nu_{\rm ff}^{\rm fill}$ (GHz) $^b$ & $\ge 0.59$   & $\ge 12$     & $\ge 3000$                  & \nodata\\
$f_{\rm fill}$                      & $\le 11\%$   & $\le 1.5\%$  & $\ge 5.8 \times 10^{-4}\%$  & \nodata\\
\hline
\multicolumn{5}{c}{Turbulence-supported WIM ($\sigma = 50\ \kms$)}\\
\hline
$n_{\rm min}$ ($\cm^{-3}$) $^c$     & $8.8$       & $120$        & $46000$       & \nodata\\
$\mean{n_{\rm min}}_V$ ($\cm^{-3}$) $^d$ & $2.3$       & $37$         & $15000$       & \nodata\\
$\nu_{\rm ff}^{\rm fill}$ (GHz) $^b$ & $\ge 0.022$ & $\ge 0.43$   & $\ge 110$     & \nodata\\
$f_{\rm fill}$                      & $\le 800\%$ & $\le 110\%$  & $\le 0.043\%$ & \nodata\\
\hline
\end{tabular}
\\$^a$: Minimum hydrogen atom density ($P_{\rm ISM} / (2 k T)$) needed to support the WIM thermally against the ISM pressure, for a temperature $T = 10^4\ \Kelv$.
\\$^b$: If the starburst is completely filled with warm ionized gas with density $n_{\rm min}$ (for thermally supported WIM) or $\mean{n_{\rm min}}_V$ (for turbulently supported WIM), $\nu_{\rm ff}^{\rm fill}$ is the frequency at which the free-free absorption optical depth over $R_{\rm SB}$ is 1.
\\$^c$: Minimum hydrogen atom density ($2 P_{\rm ISM} / (m_H \sigma^2)$) needed to support the WIM turbulently against the ISM pressure.
\\$^d$: In a turbulently supported medium with average density $n_{\rm min}$ and Mach number ${\cal M}$, $\mean{n_{\rm min}}_V$ is the median density in the volume, $n_{\rm min} / \sqrt{1 + {\cal M}^2}$.  For the Mach number, I assume a temperature $10^4\ \Kelv$.  
\\REFERENCES -- (1) \citet{YusefZadeh09} and \citet{Crocker11-Wild} for the Galactic Centre CMZ, from the IR luminosity \citet{Sanders03} combined with the \citet{Kennicutt98} IR to SFR conversion for M82, and \citet{Downes98} and \citet{Sakamoto08} for Arp 220's nuclei.\\
(2) \citet{Crocker11-Wild} for the Galactic Centre CMZ, \citet{Goetz90} and \citet{Williams10} for M82, and \citet{Sakamoto08} for Arp 220's nuclei. \\
(3) For the Galactic Centre CMZ, the listed value is a conservative estimate of the approximate magnetic and turbulent pressure from Figure 4 in \citealt{Crocker10}.  It is also roughly the thermal H II region pressure found by \citet{Zhao93}, and the \citet{Chevalier85} superwind pressure for that $\Sigma_{\rm SFR}$.  For M82, I use the hot superwind pressure from \citet{Strickland07}.  For Arp 220's radio nuclei, I take the turbulent energy density $\rho \sigma^2 / 2$, with $\sigma = 100\ \kms$ \citep{Downes98}. \\
(4) I use a scale height of 42 pc for the Galactic Centre CMZ \citep{Crocker11-Wild} and 50 pc for M82 and Arp 220's nuclei.\\
(5) \citet{Crocker11-Wild} for the Galactic Centre CMZ, \citet{Weiss01} for M82, and \citet{Downes98} for Arp 220's nuclei.
\end{minipage}
\end{table*}

There are two ways around these constraints.  First, the WIM might be supported by supersonic turbulence.  Then its density could be very small.  Interestingly, \citet{Smith06} infer a large turbulent pressure for the H II region around the M82 super star cluster A-1, so supersonic turbulence is not far-fetched.  Furthermore, the Milky Way WIM is known to be transonic or weakly supersonic \citep{Hill08,Gaensler11}.  But simulations show that large density inhomogeneities are a characteristic of mediums with supersonic turbulence.  For a Mach number ${\cal M}$, half of the volume has a density of $\la \mean{n_H} / \sqrt{1 + b^2 {\cal M}^2}$, whereas half of the mass is in clumps with density $\ga \mean{n_H} \sqrt{1 + b^2 {\cal M}^2}$, where $b \approx 1/3$ -- $1$ (\citealt*{Padoan97}; \citealt*{Ostriker01}; I use a value of $b = 1$ in this work).  Thus the uniform slab model is still formally incorrect for the WIM.

While a volume-filling WIM of arbitrarily low densities can be supported as long as the turbulent velocities are high enough, there are interesting constraints if the turbulent speeds are similar to those in the molecular medium, roughly 50 $\kms$.\footnote{\bfnop This speed is similar to the turbulent speeds of $45\ \kms$ in the M82 A-1 H II region, according to \citet{Smith06}.}  The first constraint the WIM faces is that it cannot be so dense that it free-free absorbs the $\ge \GHz$ radio emission.  In Table~\ref{table:StarburstWIMConstraints}, I calculate the free-free turnover for a volume-filling WIM that is dense enough to be in pressure equilibrium with the surrounding material.  For these results, I conservatively use the (rarefied) median density in the volume $\mean{n_{\rm min}}_V$.  While the free-free turnovers are not yet constraining for the Galactic Centre CMZ or M82, a volume-filling WIM in Arp 220's nuclei would cut off the radio emission below 110 GHz, in direct conflict with radio observations \citep{Downes98}.  The other constraint is that $f_{\rm fill}$ cannot be much smaller than 1.  Assuming the ionizing luminosity is given by equation~\ref{eqn:QwSFR}, this is again not a problem for the Galactic Centre CMZ or M82.  Note that the small amounts of free-free emission in M82 I found in section~\ref{sec:A12Results} indicate much smaller $f_{\rm fill}$.  I find that $f_{\rm fill}$ can only be a fraction of a percent in Arp 220's nuclei.  Upping the turbulent WIM speed to $100\ \kms$ in Arp 220, similar to the molecular gas turbulent speeds in that intense starburst, yields a lower average density ($\sim 13000\ \cm^{-3}$) but does not relieve these two constraints.  Much higher speeds would unbind the WIM entirely.  In short, I conclude that while a volume-filling turbulent WIM is conceivable in M82 and weaker starbursts, it is very unlikely in the most extreme ULIRG starbursts like Arp 220. 

Note at this point that the densities and pressures derived for the `WIM' in Table~\ref{table:StarburstWIMConstraints} are similar to those known to hold in H II regions in starbursts.  In the Milky Way and other normal galaxies, H II regions are overpressured, overdense regions that expand into the surrounding ISM, whereas the WIM has a pressure closer to equipartition with the rest of the ISM.  But in starbursts, the H II regions themselves have densities and pressures comparable to the surrounding ISM.  It is plausible that the H II regions \emph{are} the WIM of starbursts. 

In principle, another source of nonthermal pressure could also support the WIM.  Magnetic fields are probably high in starbursts compared to the Milky Way \citep[e.g.,][]{Thompson06}, but theory of the diffuse synchrotron emission indicates that they are either comparable to or weaker than turbulent pressure support \citep[e.g.,][]{Lacki10-FRC1}.  CRs could also provide nonthermal pressure, but current theory suggests that they are either rapidly destroyed by radiative losses in starburst environments or blown away in starburst winds, preventing them from accumulating to pressures greater than the turbulent pressure \citep[e.g.,][]{Lacki10-FRC1}.

The other way around the pressure constraint is if the WIM is a transient feature far from pressure equilibrium.  Starburst molecular media have turbulent Mach numbers of $\sim 100$, and thus have vast density contrasts.  If $n_H \approx \mean{n_H} / 100$ gas fills most of the volume of the molecular medium and becomes ionized, it forms a temporary coronal WIM.  This `WIM' would recombine within one eddy crossing time, but would be replenished as new material fills most of the volume.  In Arp 220's nuclei, a volume-filling, coronal WIM with a density $\sim 100\ \cm^{-3}$ would become opaque at $\sim 750\ \MHz$.  The main question is whether the ionizing photons can actually reach all of the low density material.  It is known that ionizing photons can travel further in turbulent media, by propagating in underdense regions, and it is thought that this is how ionizing photons escape through the Milky Way (e.g., \citealt{Ciardi02,Wood10}).  However, the typical column depth through a turbulent medium is roughly the mean column depth \citep[e.g.,][]{Ostriker01}, implying that ionizing photons would be stopped quickly in starbursts.  Also note that theoretical models of the density distribution of turbulent gas assume isothermal gas, but ionized `voids' would be much hotter, possibly invalidating those results.  A study of these issues is worthwhile.

Finally, we can consider radio observations of starburst regions.  The Galactic Centre CMZ is easily resolved, even at low frequency, and has been observed by the VLA at 74 MHz.  The H II regions in the area and on the sightline are noticeable shadows on these images.  Yet the radio synchrotron emission from this region still shines from behind the H II regions \citep{Brogan03,Nord06}.  There appears to be no volume-filling WIM in the Galactic Centre region that is opaque at 74 MHz.  Extragalactic starburst regions are much harder to resolve, but M82's starburst has been resolved at 408 MHz with MERLIN \citep{Wills97}.  The free-free absorption visible in that image is concentrated in patches.  Furthermore, some but not all of the supernova remnant radio spectra show signs of free-free absorption \citep{Wills97}, which is consistent with clumpy ionizied gas.}

\subsection{The WIM in the starburst wind}
\label{sec:WindWIM}
{\bfnop 
While the starburst region proper may be clear of warm ionized matter, H$\alpha$ images of these galaxies depict spectacular eruptions of warm gas in the winds flowing out of the starburst.  In the \citet{Chevalier85} model of starburst winds, the pressure drops rapidly past the sonic point, located roughly at the boundary of the starburst region, so rarefied warm and cold material can survive beyond that radius.  Warm gas could screen not only the starburst region, but the surrounding radio haloes of starbursts at low frequency.

We can estimate the mean density of the warm material by using the equation of continuity: $\mean{n_{\rm warm}} = \dot{M_{\rm warm}} / (m_H A v)$, where $A$ is the area of the wind.  The mass outflow of the warm material can be parametrized with a mass-loading factor, $\beta_{\rm warm} = \dot{M_{\rm warm}} / {\rm SFR}$.  While there are several models of starburst winds, they generally indicate that $\beta_{\rm warm} \approx 1$ \citep[e.g.,][]{Strickland09,Murray11}.  For example, \citet*{Hopkins12} find a mass-loading factor of $\sim 3$ for M82-like starbursts, with roughly a third of that in warm material \citep{Hopkins13}.  The density of the warm material at $R_{\rm SB}$ is roughly
\begin{align}
\nonumber \mean{n_{\rm warm}} & \approx \frac{\beta_{\rm warm} \Sigma_{\rm SFR}}{2 v m_H}\\
                     & \approx 3 \beta_{\rm warm} \ \cm^{-3} \left(\frac{\Sigma_{\rm SFR}}{50\ \Msun\ \yr^{-1}\ \kpc^{-2}}\right) \left(\frac{v}{300\ \kms}\right)^{-1}.
\end{align}
This is roughly the mean density that \citet{Shopbell98} derive from H$\alpha$ observations of M82's warm filaments.  At distances beyond $R_{\rm SB}$, the mean density drops off as $r^{-2}$, and the free-free absorption coefficient drops rapidly as $r^{-4}$.  Thus, the free-free optical depth for a face-on starburst is roughly $\alpha_{\rm H\ II} \times R_{\rm SB}$, and the turnover frequency is
\begin{eqnarray}
\nonumber \nu_{\rm ff} & \approx & 39 \beta_{\rm warm}\ \MHz\ \left(\frac{\Sigma_{\rm SFR}}{50\ \Msun\ \yr^{-1}\ \kpc^{-2}}\right) \left(\frac{R_{\rm SB}}{250\ \pc}\right)^{1/2} \\
& & \times \left(\frac{v}{300\ \kms}\right)^{-1} \left(\frac{T}{10^4\ \Kelv}\right)^{-3/4} \left(\frac{\bar{g_{\rm ff}}}{10}\right)^{1/2}.
\end{eqnarray}
While not enough to interfere at 100 MHz in M82, the wind could pose a formidable screen for Arp 220, where it would be optically thick to $\sim 2\ \GHz$ (Table~\ref{table:StarburstAbsorption}).

Starburst winds have a biconical geometry, erupting out of the starburst midplane.  It is not clear that this warm gas is actually along the line of sight along the midplane.  Edge-on starbursts like M82 may not be screened by the warm material in the wind, then.

Yet the warm material in starburst winds is not volume-filling, but is strung on filaments, both in observed starbursts and in simulations \citep[e.g.,][]{Cooper08,Cooper09,Hopkins13}.  The clumping would allow low frequency radio waves out on some sightlines, while blocking higher frequency waves on other sightlines, much like H II regions in the starburst region.   However, the amount of clumping and its distribution is not well known.  A calculation of the free-free absorption within the wind would be useful.}

\subsection{Free-free absorption in hot wind plasma}
\label{sec:StarburstHotWind}
The high rate of supernovae in starburst galaxies occurring within a relatively small space is expected to excavate a hot phase of the ISM (\citealt{McKee77}; \citealt{Heckman90}; \citealt{Lord96}).  The hot ISM erupts as a starburst-wide superwind, and is low density, but high in pressure, temperature, and, it is thought, filling factor \citep[e.g.,][]{Chevalier85}.  Gas emitting in soft X-rays is indeed observed in many starbursts, though \citet{Strickland00} argue that this emission comes from a cooler phase with lower filling factor than the actual wind.  In addition, \emph{Chandra} detected diffuse 6.7 keV iron line emission in M82 that supports the presence of $10^8\ \Kelv$ plasma \citep{Strickland07}.  

\citet{Strickland09} give the central density of the superwind {\bfnop from a thin disk} as
\begin{align}
\nonumber n_c & = {\bfnop 1.860 \frac{\beta_{\rm hot}^{3/2}}{\epsilon_{\rm therm}^{1/2}} \frac{\dot{M}^{3/2}}{\dot{E}^{1/2} m_H (\pi R_{\rm SB}^2)}}\\
 & = {\bfnop 0.93\ \cm^{-3} \left(\frac{\beta_{\rm hot}}{2}\right)^{3/2} \left(\frac{\epsilon_{\rm therm}}{0.75}\right)^{-1/2} \left(\frac{\Sigma_{\rm SFR}}{50\ \Msun\ \yr^{-1}\ \kpc^{-2}}\right)}
\end{align}
and the central temperature for a completely ionized {\bfnop hydrogen} wind as
\begin{equation}
{\bfnop T_c = \frac{0.2 m_H \epsilon_{\rm therm} \dot{E}}{k \beta_{\rm hot} \dot{M}} = 3.0 \times 10^7\ \Kelv \left(\frac{\epsilon_{\rm therm}}{\bfnop 0.75}\right) \left(\frac{\beta}{2}\right)^{-1}.}
\end{equation}
In these equations, $\epsilon_{\rm therm} \approx {\bfnop 0.75}$ is the fraction of supernova {\bfnop mechanical} energy that ends up in thermal energy of the plasma, $\beta_{\rm hot} \approx 2$ is the mass-loading fraction\footnote{{\bfnop The factor $\beta_{\rm hot}$ is the} fraction of mass ejected by stellar winds and supernova that ends up in the wind; it can be greater than 1 if cold gas is swept up by the wind as it leaves the galaxy.  {\bfnop I converted the SFR above $1\ \Msun$ in \citet{Strickland09} to the SFR for a Salpeter IMF from 0.1 to 100 $\Msun$ using ${\rm SFR}(\ge 1\ \Msun) = 0.45\ {\rm SFR}$.}} \citep{Strickland09}.  

I find that the high temperatures and low densities of starburst winds makes them extremely poor at free-free absorption.  {\bfnop For $\bar{g_{\rm ff}} = 20$,} the turnover frequency is far below observability:
\begin{equation}
\nu_{\rm ff} \approx {\bfnop 24}\ \kHz\ \left(\frac{\Sigma_{\rm SFR}}{50\ \Msun\ \yr^{-1}\ \kpc^{-2}}\right) \left(\frac{s}{100\ \pc}\right)^{1/2}.
\end{equation}
I list these $\nu_{\rm ff}$ in Table~\ref{table:StarburstAbsorption} for representative starbursts.  The starburst wind introduces free-free absorption only at frequencies less than {\bfnop $\sim 1$ -- 2000} kHz, and is not important at the observable MHz radio frequencies fit by models.

\subsection{Free-free absorption from cosmic ray-ionized molecular gas}
\citet{Thompson05} have argued that cold molecular gas, instead of rarefied supernova-heated material, fills most of the volume of starbursts.  This could happen if supernova remnants rapidly lose their kinetic energy to radiative losses as they expand in a dense molecular medium, so that supernova-heated material fills only small isolated bubbles {\bfnop \citep{Thornton98}}.  Then the molecular gas {\bfnop fills} most of the starburst region, since the H II regions also have a small filling factor.  Molecular gas is mostly neutral and dust extinction rapidly {\bfnop extinguishes} any ultraviolet light {\bfnop (but see the caveats in section~\ref{sec:StarburstWIM})}.  However, cosmic rays should provide a relatively high level of ionization through fairly large columns (\citealt*{Suchkov93}; \citealt{Papadopoulos10-CRDRs}), though the details of CR diffusion in starbursts and their penetration into molecular {\bfnop clouds} is poorly understood.  

As I argued in \citet{Lacki12-GRDRs}, the {\bfnop CR} ionization rate in starburst galaxies, where the proton spectrum is relatively hard, is 
\begin{align}
\zeta_{\rm CR} & = \frac{\eta_{\rm ion} L_{\rm CR} m_H}{E_{\rm ion} M_H}\\
& {\bfnop = 1.8 \times 10^{-15}\ \sec^{-1} \left(\frac{\rm \tau_{\rm gas}}{20\ \Myr}\right)^{-1} \left(\frac{\eta_{\rm ion}}{0.1}\right).}
\end{align}
In this equation, $L_{\rm CR}$ is the luminosity of injected cosmic rays, $E_{\rm ion} \approx 30\ \eV$ is the energy lost per cosmic ray ionization event \citep{Cravens78}, $M_H$ is the mass of gas in the galaxy, and $\eta_{\rm ion} \approx 0.1$ is the fraction of cosmic ray power that goes into ionization.  {\bfnop In the Galactic Centre CMZ, winds probably remove CRs before they can ionize material, so $\eta_{\rm ion}$ is much lower than 0.1 \citep{Crocker11-Wild}.}  The injected cosmic ray luminosity is thought to be approximately $10^{50}\ \erg$ per supernova, {\bfnop and the} supernova rate is $\Gamma_{\rm SN} = 0.0064\ \yr^{-1} ({\rm SFR} / \Msun\ \yr^{-1})$, {\bfnop for a Salpeter IMF extending from 0.1 to $100\ \Msun$.  I have combined the SFR and $M_H$ into a single parameter $\tau_{\rm gas} \equiv M_H / {\rm SFR}$, which is about $\sim 20\ \Myr$ in nuclear starbursts.}

The ionization fraction is set by the ratio of the gas density $n_H$ and a characteristic density $n_{\rm ch} \approx 1000\ (\zeta_{\rm CR} / 10^{-17}\ \sec^{-1})\ \cm^{-3}$  from \citet{McKee89}.  In the cosmic ray ionized gas of starbursts, $n_{\rm ch} \gg n_H$ and $x_e \approx 10^{-7} (n_{\rm ch} / n_H)$.  Thus,
\begin{equation}
x_e \approx {\bfnop 10^{-5}\ \left(\frac{\zeta_{\rm CR}}{10^{-15}\ \sec}\right) \left(\frac{n_H}{1000\ \cm^{-3}}\right)^{-1}}
\end{equation}

While the ionization fraction is low, free-free absorption is enhanced by two factors: molecular material is dense, so the density of electrons and ions is relatively high, and molecular gas is cold.  The electrons and ions reach thermal equilibrium with the surrounding gas long before they recombine \citep[e.g.,][]{McCall02}.  Assuming typical starburst molecular gas temperatures of $\sim 100\ \Kelv$ and that $n_e = n_i {\bfnop =} x_e n_H$ and $\bar{g_{\rm ff}} = 10$, the frequency of the free-free spectrum turnover is 
\begin{equation}
{\bfnop
\nu_{\rm ff} \approx 2.4\ \MHz\ \left(\frac{\zeta_{\rm CR}}{10^{-15}\ \sec^{-1}}\right) \left(\frac{T}{100\ \Kelv}\right)^{-3/4} \left(\frac{s}{100\ \pc}\right)^{1/2}}
\end{equation}
{\bfnop Note that there is no density dependence, so inhomogeneities do not affect the free-free absorption turnover unless $\zeta_{\rm CR}$ itself varies.}

\subsection{The Razin effect}
\label{sec:Razin}
Free-free absorption is not the only process that cuts off the radio spectrum at low frequency.  At low energies, the index of refraction of plasma suppresses the beaming of synchrotron radiation and causes it to fall off exponentially \citep{Rybicki79}.  The frequency where this Razin effect becomes important is
\begin{equation}
\nu_R = 185\ \kHz \left(\frac{n_e}{1\ \cm^{-3}}\right) \left(\frac{B}{100\ \muGauss}\right)^{-1},
\end{equation}
from \citet{Schlickeiser02}.

Suppose all the gas in the starburst, both ionized and neutral, has the same magnetic field.  $B$ and $\Sigma_{\rm SFR}$ are likely to depend on each other.  The existence of the linear far-infrared radio correlation of galaxies constrains magnetic field strengths to be 
\begin{equation}
\label{eqn:BFRC}
B_{\rm FRC} \approx 470\ \muGauss \left(\frac{\Sigma_{\rm SFR}}{50\ \Msun\ \yr^{-1}\ \kpc^{-2}}\right)^{1/2}
\end{equation}
from \citet{Lacki10-FRC1}, after using the \citet{Kennicutt98} Schmidt law to convert between gas surface density and star-formation surface density.  The Razin cutoff in the hot superwind {\bfnop of starbursts on the FIR-radio correlation} is
\begin{equation}
{\bfnop
\nu_R^{\rm hot} = 37\ \kHz \left(\frac{\Sigma_{\rm SFR}}{50\ \Msun\ \yr^{-1}\ \kpc^{-2}}\right)^{1/2},
}
\end{equation}
varying between {\bfnop 5 and 500 kHz} for starbursts with $\Sigma_{\rm SFR}$ between 1 and $10^4\ \Msun\ \yr^{-1}\ \kpc^{-2}$.  {\bfnop Table~\ref{table:StarburstAbsorption} lists the superwind Razin frequencies in a few other starburst regions}.

If starbursts are instead filled with cosmic ray ionized gas, the low electron density ($n_e = x_e n_H$) of these regions imply even lower Razin cutoffs:
\begin{equation}
{\bfnop
\nu_R^{\rm mol} = \displaystyle 0.39\ \kHz \left(\frac{\zeta_{\rm CR}}{10^{-15}\ \sec^{-1}}\right) \left(\frac{\Sigma_{\rm SFR}}{50\ \Msun\ \yr^{-1}\ \kpc^{-2}}\right)^{-1/2}
}
\end{equation}
In high density molecular regions, Zeeman splitting measurements indicate the magnetic field may be even higher \citep*{Robishaw08}, so the Razin effect could be even less important.  

{\bfnop Finally,} in H II regions, which are fully ionized and high density:
\begin{equation}
\nu_R^{\rm H II} = \displaystyle 39\ \MHz\ \left(\frac{n_H}{1000\ \cm^{-3}}\right) \left(\frac{\Sigma_{\rm SFR}}{50\ \Msun\ \yr^{-1}\ \kpc^{-2}}\right)^{-1/2}
\end{equation}
Within these regions, we see that free-free absorption -- which turns H II regions opaque at GHz frequencies -- is more important than the Razin effect.

I conclude that the Razin effect is not observable in starburst galaxies.

\subsection{Synchrotron self-absorption}
Both free-free absorption and {\bfnop the} Razin cutoff depend on the density distribution of ionized matter, something {\bfnop that} is {\bfnop uncertain} in starburst galaxies.  However, because we observe synchrotron emission from starburst galaxies, there must also be synchrotron self-absorption in them as well.

The maximum brightness temperature of a synchrotron source {\bfnop is} limited by synchrotron self-absorption {\bfnop to}
\begin{equation}
\label{eqn:TSynchMax}
T_{\rm max} \approx 3 \times 10^{12}\ \Kelv\ \left(\frac{\nu}{\GHz}\right)^{1/2} \left(\frac{B}{100\ \muGauss}\right)^{-1/2}
\end{equation}
from \citet*{Begelman84}.  {\bfnop While the Razin effect can suppress synchrotron absorption \citep{Crusius88}, I established it is not important for starbursts in section~\ref{sec:Razin}.}

For resolved starbursts with an observed low frequency radio spectrum, it is possible to simply fit a model to the radio spectrum and see when, if ever, the brightness temperature $T_b$ is greater than $T_{\rm max}$.  In general, though we expect starbursts to lie on the far-infrared radio correlation, with $\nu L_{\nu} (1\ \GHz) = 10^{-6} L_{\rm TIR} = 5630\ \Lsun ({\rm SFR}/\Msun\ \yr^{-1})$ (\citealt{Kennicutt98}; \citealt*{Yun01}).  Ignoring geometrical factors, the starburst can be thought of as a sphere with radius $R_{\rm SB}$, so that the {\bfnop brightness temperature} is $T_b = L_{\nu} c^2 / (8\pi^2 k \nu^2 R_{\rm SB}^2)$ {\bfnop in the Rayleigh-Jeans limit}.  Likewise, the star-formation surface density is $\Sigma_{\rm SFR} = {\rm SFR} / (\pi R_{\rm SB}^2)$.  Putting all {\bfnop of} these formulas together, I find that the FIR-radio correlation implies
\begin{equation}
\label{eqn:Tb1GHzFRC}
{\bfnop
T_b (1\ \GHz) \approx 30000\ \Kelv \left(\frac{\Sigma_{\rm SFR}}{50\ \Msun\ \yr^{-1}\ \kpc^{-2}}\right).
}
\end{equation}

Now we need to extrapolate this down to low radio frequencies.  The simplest assumption is that the radio spectral index is constant, and {\bfnop the brightness temperature has the form} $T_b = T_b (1\ \GHz) (\nu / \GHz)^{-2.7}$. {\bfnop Then I can use} eqn.~\ref{eqn:BFRC} for $B$, and {\bfnop by} equating eqns.~\ref{eqn:TSynchMax} and~\ref{eqn:Tb1GHzFRC}, I find the synchrotron self-absorption turnover is at:
\begin{equation}
{\bfnop
\nu_{\rm SSA} = 2.4\ \MHz\ \left(\frac{\Sigma_{\rm SFR}}{50\ \Msun\ \yr^{-1}\ \kpc^{-2}}\right)^{0.39}.
}
\end{equation}
For starbursts on the FIR-radio correlation with $1 \le \Sigma_{\rm SFR}/(\Msun\ \yr^{-1}\ \kpc^{-2}) \le 10^4$, $\nu_{\rm SSA} \approx 0.9 - 30\ \MHz$.  This result is conservative {\bfnop in the sense that the actual synchrotron self-absorption is weaker, because} the radio spectra {\bfnop are free-free absorbed and} tend to be flatter than $\nu^{-0.7}$.  {\bfnop I list the approximate synchrotron self-absorption frequency of the Galactic Centre CMZ, M82's starburst, and Arp 220's radio nuclei in Table~\ref{table:StarburstAbsorption}.}

\subsection{Summary}

\begin{table*}
\begin{minipage}{170mm}
\caption{Low frequency absorption in starbursts}
\label{table:StarburstAbsorption}
\begin{tabular}{lcccc}
\hline
Property                                          & Galactic Centre CMZ & M82 & Arp 220 Nuclei & References\\
\hline
${\rm SFR}$ ($\Msun\ \yr^{-1}$)                   & 0.07                 & 10              & 100        & (1)\\
$R$ ($\pc$)                                       & 112                  & 250             & 100        & (1)\\
$B$ ($\muGauss$)                                  & 75			 & 200		   & 6000       & (2)\\
$S_{\nu} (1.4\ \GHz)$ (Jy)                        & 1915                 & 5.5             & $\sim 0.1$ & (3)\\
$T_b (1.4\ \GHz)$ ($\Kelv$)                       & 50                   & 6000            & 340000     & (4)\\
$\Sigma_{\rm SFR}$ ($\Msun\ \yr^{-1}\ \kpc^{-2}$) & 2                    & 50              & 3000       & \nodata\\
$\mean{n_H}$ ($\cm^{-3}$)                         & 120                  & 410             & 10000      & (1)\\
$\nu_{\rm ff}$ (warm wind; MHz)                   & 1.0                  & 39              & 1500       & \nodata\\
$\nu_{\rm ff}$ (hot wind; MHz)                    & 0.0010               & 0.038           & 1.5        & \nodata\\
$\nu_{\rm ff}$ (molecular; MHz)                   & $\ll 4$              & $\la 4$         & $\sim 4.3$ & \nodata\\
$\nu_R$ (hot wind; MHz)                           & 0.0092               & 0.085           & 0.17       & \nodata\\
$\nu_R$ (molecular; MHz)                          & $\ll 5 \times 10^{-4}$ & $\la 7 \times 10^{-4}$ & $5.5 \times 10^{-4}$ & \nodata\\
$\nu_{\rm SSA}$ (${\cal B} = -0.7$; $\Psi_{\rm SB} = 1$; MHz) & 0.68                 & 3.4             & 19         & \nodata\\
\hline
\end{tabular}
\\References -- (1) See Table~\ref{table:StarburstWIMConstraints}.\\
(2) \citet{Crocker11-Wild} for the Galactic Centre CMZ, \citet{deCeaDelPozo09-M82} and \citet{Persic08} for M82, and \citet{Torres04-Arp220} for Arp 220's nuclei.  \\
(3) \citet*{Reich90} (via \citet{Crocker11-Wild}) for the Galactic Centre CMZ, \citet{Adebahr12} for M82, and \citet{Downes98} for Arp 220's nuclei.\\
(4) I use distances of 8.0 kpc for the Galactic Centre, 3.6 Mpc for M82, and 80 Mpc for Arp 220.\\
\end{minipage}
\end{table*}

Of the processes I considered, the most likely to suppress low frequency radio emission from starbursts {\bfnop is WIM in the starburst wind beyond the starburst region proper.  This material is inhomogeneous, so its effects are unknown.  I show that a volume-filling WIM is unlikely to exist within starburst regions themselves.  Besides WIM in the starburst wind, the starburst's H II regions dominate absorption down to $\sim 10\ \MHz$.  Internal processes within the starburst are generally unimportant, with a maximum cutoff frequency of $\sim 20\ \MHz$ for synchrotron self-absorption in Arp 220-like starbursts.  Free-free absorption from the volume-filling phases is unimportant down to a few MHz, as is the Razin effect.}

\section{Conclusions}
With the renewed interest in low frequency radio astronomy, it is time for a better theoretical understanding of the low frequency radio spectra of starburst galaxies.  

Previous models of the emission at these frequencies, if they considered free-free absorption at all, predicted that starbursts are opaque below a GHz.  This was because they used the uniform slab model, assuming that ionized gas evenly pervaded the starburst.  Most of the ionized gas mass is not truly diffuse in starbursts, but instead resides in discrete H II regions.  If the H II regions are uniformly distributed throughout the starburst, they contribute to an effective absorption coefficient which should be used in the uniform slab formula.  This coefficient does not approach infinity at low frequencies (Figures \ref{fig:AlphaEffWNuIndiv} -- \ref{fig:AlphaEffWNuAC}).  {\bfnop If} H II regions partially cover the starburst, {\bfnop they leave} sightlines where the emission is unobscured.  Furthermore, the H II regions are usually some way in from the surface of the starburst, so there is unobscured synchrotron emitting material on a sightline in front of the nearest H II region.  

{\bfnop To demonstrate this method, I fit the radio spectrum of M82 with the effective absorption coefficient} in the limit that the H II regions do not fill most of the starburst.  The calculation ultimately reduces down to the density of the H II regions multiplied by the cross section of each to absorb background radio waves.  I applied the calculations to the radio spectrum of M82.  I find that models with discrete H II regions, around either individual O{\bfnop B} stars or Super Star Clusters, are able to reproduce the radio spectrum reasonably well (see Figures~\ref{fig:FitM82RadioSpectrumIndiv}--\ref{fig:FitM82RadioSpectrumAC}).  

The models I presented in the paper were relatively simple, assuming that H II regions are characterized by a single temperature and density.  However, if the temperature and density distribution is known through other means, such as radio recombination lines, it is straightforward to integrate up the cross sections and calculate the effective absorption coefficient.  A more important flaw in the approach described here is that I assume that H II regions are uniformly dense Str\"omgren spheres.  However, dust absorption affects both the amount of ionizing photons available to produce H II regions \citep{Petrosian72}, and the structure of H II regions \citep{Draine11-DustyHII}.  Since my models suggest a low amount of free-free emission, which could mean that dust is absorbing ionizing photons, it is important that these effects be studied.  

{\bfnop The other big factor that I have neglected is turbulence in H II regions, which is suggested by the observations of \citet{Smith06}.  Supersonic turbulence introduces large density fluctuations, invalidating the assumption of uniform density within the Str\"omgren spheres.  The molecular gas in starbursts is known to be highly supersonic \citep[e.g.,][]{Downes98}. \citet{Mellema06} ran simulations of H II regions expanding through a turbulent ISM, finding the resulting H II regions are neither smooth nor uniformly ionized.  A simulation of the free-free absorption from turbulent H II regions would be helpful, although there may be no easy formula for describing the radio absorption with a turbulent ISM.}

This work implies that starbursts are in fact fairly bright at low frequencies.  I considered whether any other absorption process could cause a turnover in the radio spectrum.  {\bfnop While the are several factors inhibiting WIM growth in the starburst region proper, the warm gas in the starburst wind may prove to be a screen at low frequency, although those conclusions need to be verified.  Internal processes within the starburst region -- free-free absorption from the volume-filling gas phases, the Razin effect, and synchrotron self-absorption -- are not important except below a few MHz.  Thus, the free-free absorption from discrete H II regions is the essential ingredient for describing the radio spectrum at observable frequencies.}

\section*{Note Added In Proof}
Equation 4 is actually appropriate for a Salpeter IMF from 1 - 100 $M_{\odot}$, not 0.1 - 100 $M_{\odot}$ as stated in the text.  This error affects the `Simple Sources' and `Simple SSCs' models, in Sections 2 and 3 and Figures 1 and 3.  The values are right if all SFRs and stellar masses are assumed to include only stars with mass greater than 1 $M_{\odot}$.  To convert $\alpha_{\rm eff}$ to 0.1 - 100 $M_{\odot}$, multiply it by 0.39.  Likewise, the derived SFRs for these models of M82 can be divided by 0.39 to get the SFR between 0.1 and 100 $M_{\odot}$.  The `Aging Clusters' models were already converted to the correct IMF.

\section*{Acknowledgments}
{\bfnop I} was supported by a Jansky Fellowship from the National Radio Astronomy Observatory.  The National Radio Astronomy Observatory is operated by Associated Universities, Inc., under cooperative agreement with the National Science Foundation.  I am grateful to Rainer Beck and Diego Torres for comments on this research.

\clearpage
\begin{appendix}
\section{A Method for Calculating Free-Free Absorption from Discrete H II Regions}
\label{sec:Derivation}
\subsection{Derivation of the effective absorption coefficient}
\begin{figure}
\centerline{\includegraphics[width=8cm]{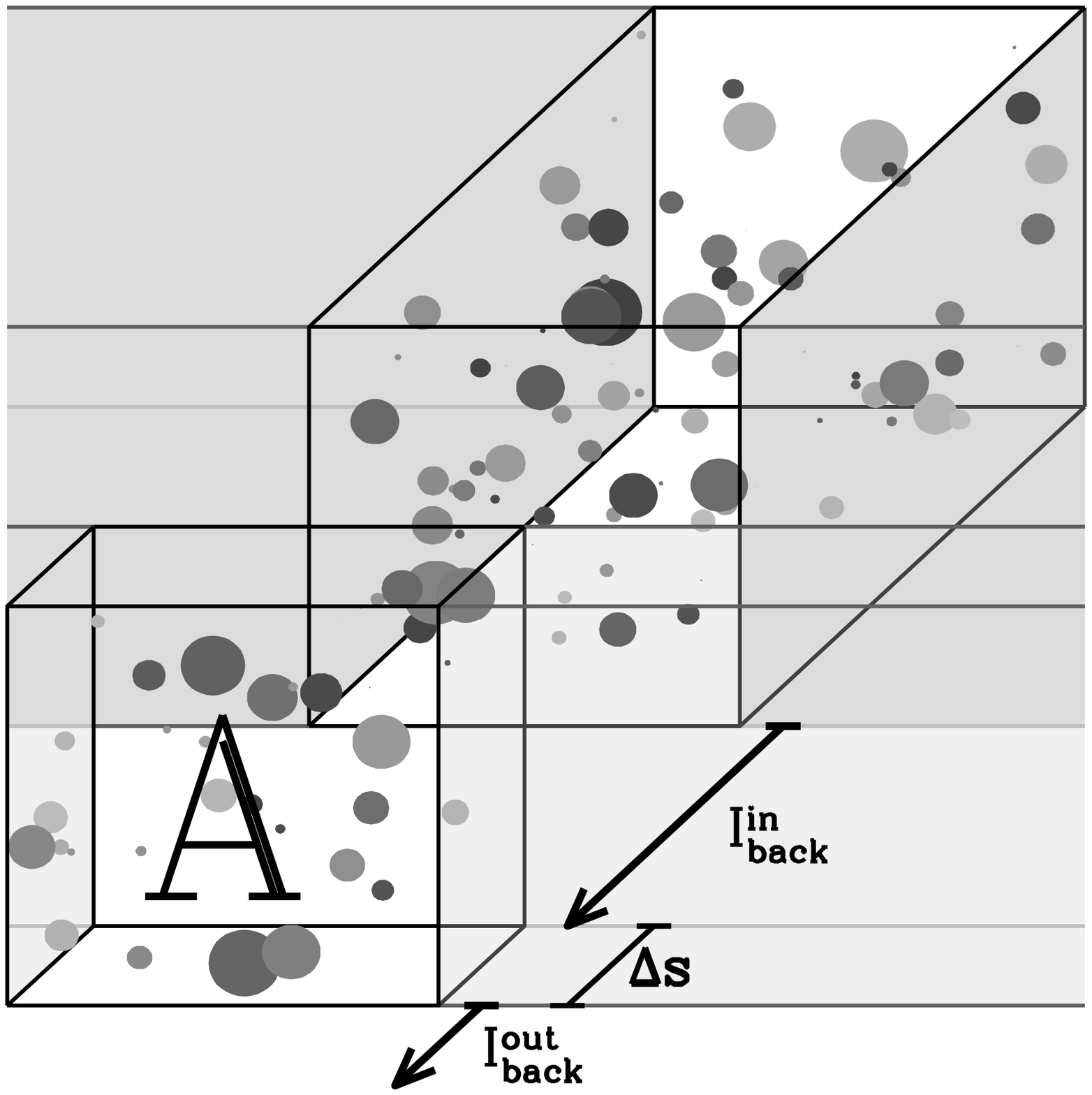}}
\caption{{\bfnop A sketch of how} H II regions (filled circles) in a slice of the starburst with area $A$ and thickness $\Delta s$ absorb radiation from the column behind it.  {\bfnop Different shades of grey in the H II regions can represent differing physical properties like electron density or temperature.}  The column can be small relative to the rest of the starburst (shaded in grey), but each slice is large enough to contain a representative sample of H II regions.  A beam of light is emitted from the background column, enters perpendicular to the slice with intensity $I_{\rm back}^{\rm in}$ and emerges with intensity $I_{\rm back}^{\rm out}$, where $I$ is flux per solid angle.  The fraction of light absorbed passing through the slice, which can be calculated by summing over individual H II regions in the slice, is directly related to the mean effective absorption coefficient of the starburst from H II regions.  The coefficient then can be used in a uniform slab model (as suggested by the uniform grey shading of the rest of the starburst). \label{fig:Explanation}}
\end{figure}

I start by considering how a thin slice of the starburst with thickness $\Delta s$ absorbs background radio emission passing perpendicularly through it (Figure~\ref{fig:Explanation}).  The slice has an area $A$ and thickness large enough to contain a representative sample of the starburst's H II regions, but otherwise can be small compared to the starburst.  {\bfnop I am basically assuming that there is some scale smaller than the starburst size on which the number density (or distribution function) of H II regions is well-defined.}  I ignore the H II regions' effects on radio emission emitted within the slice -- as long as the slice is thin, and as long as the H II regions have a small filling factor, this should be a valid approximation.  If the filling factor is large, a true uniform slab model is more accurate anyway.  

Suppose that there are $M$ H II regions (indexed by the number $m$).  Each H II region passes a fraction $\Phi_m$ of the normal incident radiation $I_{\rm back}^{\rm in} A$ on the slice, where $I$ is flux per solid angle.  For example, if a totally opaque sphere of radius $r$ sat in a slice of area $A$, $\Phi =  1 - (\pi r^2 / A)$.  Then the background radio intensity after passing through the slice is
\begin{equation}
I_{\rm back}^{\rm in} A \left[\prod_{m = 1}^M \Phi_m\right] = I_{\rm back}^{\rm out} A.
\end{equation}
Taking the logarithm of both sides, we can rephrase this as an optical depth:
\begin{equation}
-\sum_{m=1}^M \ln \Phi_m = -\ln\left(\frac{I_{\rm back}^{\rm out}}{I_{\rm back}^{\rm in}}\right) \equiv \Delta \tau_{\rm eff}.
\end{equation}
Because each H II fraction only covers a small portion of the starburst, a convenient approximation is to define $\phi_m = 1 - \Phi_m$ and then to assume that $\ln \Phi_m \approx -\phi_m$.  This is equivalent to saying that the H II regions do not overlap within the slice.  Then we have
\begin{equation}
\sum_{m=1}^M \phi_m \approx \Delta \tau_{\rm eff}.
\end{equation}

If there are $J$ types of H II regions, where each type of H II region absorbs the same amount of background radio, and if there are $K_j$ H II regions of each type $j$ within the slice, then 
\begin{equation}
\Delta \tau_{\rm eff} \approx \sum_{j = 1}^J K_j \phi_j.
\end{equation}
More realistically, instead of having a few distinct types of H II regions, {\bfnop we might have} a distribution function parametrizing the absorption properties of H II regions.  Suppose the number of H II regions with some relevant physical quantity (such as radius) between $q$ and $q + dq$ and within a volume $dV$ is $dN/(dq dV)$.  Then, since the volume of the slice is $A \Delta s$, the effective optical depth is
\begin{equation}
\Delta \tau_{\rm eff} \approx \int \frac{dN}{dq dV} A \Delta s \phi(q) dq.
\end{equation}
Of course, this can easily be generalized to more than one parameter.

The effective optical depth across the slice translates to an effective absorption coefficient $\alpha_{\rm eff} \equiv \Delta \tau_{\rm eff} / (\Delta s)$ of the starburst.  Furthermore, the quantity $A \phi(q)$ is an effective absorption cross section $\sigma$ for each H II region.  So we have
\begin{equation}
\nonumber \alpha_{\rm eff} \approx \int \frac{dN}{dq dV} \sigma(q) dq.
\end{equation}
Therefore, the absorption coefficient reduces to a number density times a cross section, as might be expected.

If the density of H II regions is constant throughout the starburst, the equation of radiative transfer through the starburst, $dI/ds \approx -\alpha_{\rm eff} I + j$, has the uniform slab solution:
\begin{equation}
I = \frac{j}{\alpha_{\rm eff}} \left(1 - e^{-\tau_{\rm eff}}\right)
\end{equation}
except that the optical depth is the effective optical depth defined as $\tau_{\rm eff} = \alpha_{\rm eff} s$, where $s$ is the sightline length through the starburst.  Of course, the density of H II regions may itself vary, for example, decreasing towards the edge of the starburst.  However, the basic principle remains the same: the effective absorption coefficient is calculated at each point using equation~\ref{eqn:alphaEff}, and then used in the radiative transfer equation to find the synchrotron intensity on each sightline.

\subsection{The covering fraction}
The calculation of the covering fraction of the H II regions is analogous to the calculation of the absorption of the radio flux.  Essentially, we are interested in how much of the background sky is blocked by the H II regions.  We can phrase this in terms of a flux calculation: if H II regions are totally opaque, the covering fraction is equal to the fraction of a uniform background intensity they absorb.  Thus, we have for a distribution function of H II regions
\begin{equation}
\alpha_{\rm cover} = \int \frac{dN}{dq dV} \pi R_S^2 dq
\end{equation}
if each H II region is a sphere with radius $R_S$.  

For the covering fraction calculation, the H II regions are treated as a foreground screen rather than as a uniform slab, as {\bfnop a H II region covers the background no matter how deep into the starburst it is}.  From the solution to absorption from a foreground screen, the probability any given sightline is covered is $P_{\rm cover} = 1 - e^{-\alpha_{\rm cover} s}$, where $s$ is the length of the sightline, assuming $\alpha_{\rm cover}$ is uniform throughout the starburst.

The length of a sightline may vary; for example, in an edge-on disc, it is shorter towards the rim than through the centre.  The covering fraction for the entire starburst is the average of the covering probability $P_{\rm cover}$ over all sightlines that pass through the starburst:
\begin{equation}
f_{\rm cover} = \frac{1}{\Omega_{\rm obs}} \int (1 - e^{-\alpha_{\rm cover} s}) d\Omega_{\rm obs}.
\end{equation}
In this equation, $\Omega_{\rm obs}$ is the solid angle subtended by the entire starburst from Earth.  Note that since starbursts are small on the sky, $\Omega_{\rm obs}$ is directly proportional to the projected area of the starburst{\bfnop,} $A_{\rm proj}${\bfnop .}

{\bfnop One} consequence of $f_{\rm cover} < 1$ is that we expect the optical depth to increase to some maximum value at low frequency and no further.  If there are just a few H II regions, then the covering fraction is small and parts of the starburst remain completely unobscured.  However, even if the density of H II regions is high, discrete H II regions tend to overlap from our point of view and cover each other instead of background material, {\bfnop causing $f_{\rm cover}$ to grow more slowly than $\tau_{\rm eff}$, and be slightly less than 1.  In practice, the small filling factor of H II regions is the more important effect when $\tau_{\rm eff} \gg 1$ (see section~\ref{sec:fFillCalc}).}

\subsection{The filling fraction}
\label{sec:fFillCalc}
Likewise, we can calculate the filling fraction {\bfnop for} a distribution of Str\"omgren spheres.  The starburst has a volume $V_{\rm SB}$, and a Str\"omgren sphere of radius $R_S$ has volume $4/3 \pi R_S^3$.  Therefore, each sphere leaves a fraction $1 - 4 \pi R_S^3 / (3 V_{\rm SB})$ of the starburst unfilled, and the filling factor {\bfnop is}
\begin{equation}
1 - f_{\rm fill} \approx \exp\left[-\int \frac{dN}{dq} \left(\frac{4 \pi R_S(q)^3}{3 V_{\rm SB}}\right) dq\right].
\end{equation}
When the filling factor is small, the spheres do not overlap, and 
\begin{equation}
\label{eqn:fFill}
f_{\rm fill} \approx \int \frac{dN}{dq} \left(\frac{4 \pi R_S(q)^3}{3 V_{\rm SB}}\right) dq.
\end{equation}

Since $R_S$ is much smaller than the size of the starburst, the filling factor of H II regions can be much smaller than 1 even if the covering fraction is nearly 1.  As a result, a sightline through the starburst will not immediately intercept an H II region, even if the covering fraction is nearly 1.  Instead, the first absorbing H II region on the sightline is buried past unobscured synchrotron-emitting space: this ensures that there is still synchrotron radio flux at low frequencies even if the covering fraction is high.

\subsection{Limitations of the approach}
{\bfnop While the approach I outlined above has more power than, for example, simply assuming all spheres are at the midplane and have the same radius, it still suffers some limits.  First, the cross section depends on the structure of the H II regions.  I assume that each H II regions is a uniformly dense sphere with radius $R_S$.  Actual H II regions have more complex structures \citep[e.g.,][]{Mellema06}.

On a more fundamental level, though, the approach assumes the H II regions have a well-defined number density, or local distribution function.  But the relatively small numbers of H II regions per starburst introduces error.  Some H II regions are too rare (for example, large SSCs), with perhaps one or two per starburst region, and formally invalidating the use of a local distribution function.  As the considered volumes become smaller, stochastic effects become more important. The approach breaks down completely at small scales when there is fewer than $\sim 1$ H II region per volume.

To quantify this a bit, in the simple sources model (section~\ref{sec:IndivOStars}) there are $\sim 22000$ H II regions per $1\ \Msun\ \yr^{-1}$ of star-formation.  In M82, if ${\rm SFR}_{\rm eff} \approx 1\ \Msun\ \yr^{-1}$ (section~\ref{sec:A12Results}), then there is about 1 H II region per $(10\ \pc)^3$, so the density is undefined on scales below $\sim 10\ \pc$.  Since I am considering the starburst as a whole, my approximation should work.  There are fewer H II regions in the simple SSC model (section~\ref{sec:SSCs}), 1300 per $1\ \Msun\ \yr^{-1}$ of star-formation for $M_l = 1000\ \Msun$, $M_c = 5 \times 10^6\ \Msun$, and $t_{\rm ion} = 10^7\ \yr$.  There are fewer SSCs of large mass: $dN/d\ln M_{\star} \approx 1300 (M_{\star} / 10^3\ \Msun)^{-1} ({\rm SFR}_{\rm eff}/(\Msun \yr^{-1}))$.  Thus, there is $\sim 1$  $M_{\star} \approx 10^6\ \Msun$ SSCs in M82, and the absorption from these big H II regions is very poorly calculated.  Fortunately, the $\sim 1000$ small SSCs are the source of most of the free-free absorption (see the discussion of clustering in section~\ref{sec:SSCs}), so the assumption of a distribution function is sufficient for calculating the total absorption throughout the entire starburst.  

None the less, there is need for a better quantification of stochastic effects.}

\section{Summary of the radiative properties of Str\"omgren spheres}
\label{sec:StromgrenRadiative}
\subsection{The effective cross section for translucent spheres}
\label{sec:EffectiveSigma}
I treat H II regions as Str\"omgren spheres, which absorb the radiation behind them.  The effective cross section of a sphere of radius $R_S$ and absorption coefficient $\alpha_{\rm H II}$ is equal to the projected area of the H II region times the fraction of background flux it obscures:
\begin{eqnarray}
\label{eqn:SigmaTranslucentSphere}
\nonumber \sigma & = & \int_0^{R_S} 2 \pi y \left[1 - e^{-2 \alpha_{\rm H II} \sqrt{R_S^2 - y^2}}\right] dy \\
       & = & \pi R_S^2 - \frac{\pi}{2 \alpha_{\rm H II}^{2}} \left[1 - e^{-2\alpha_{\rm H II} R_S}(1 + 2 \alpha_{\rm H II} R_S)\right]
\end{eqnarray} 
When $\alpha_{\rm H II}$ is very large (at low frequency, for example), the spheres become totally opaque and the effective cross section is just $\pi R_S^2$.  If instead $\alpha_{\rm H II}$ is very small, as at high frequency, the effective cross section goes as $V_S = (4/3) \pi \alpha_{\rm H II} R_S^3$.  

{\bfnop Note that when all of the Str\"omgren spheres are optically thin, then $\alpha_{\rm eff} \approx f_{\rm fill} \alpha_{\rm H II}$.  This is identical to the absorption coefficient in the uniform slab case when the electron density is equal to the volume-averaged electron density.  Discrete H II region models diverge from the uniform slab solution only at low frequencies when H II regions start becoming opaque.}

\subsection{Luminosity of a translucent sphere}
{\bfnop 
In order to calculate the free-free emission correctly, I must not only consider the absorption from the population of H II regions distributed throughout the starburst, but the free-free self-absorption within the H II region itself.  For example, a very dense H II region may have a free-free absorption turnover at a few GHz, even {\bfnop if} the starburst as a whole is transparent down to several hundred MHz.

The emergent free-free flux at a point on the surface of a uniform sphere is 
\begin{equation}
F_{\nu} = 2 \pi B_{\nu} (T_e) \int_0^1 [1 - e^{-2 \alpha_{\nu} R_S \mu}] \mu d\mu,
\end{equation}
where $\mu$ is the cosine of the angle between the centre of the sphere and a line of sight and $R_S$ is the radius of the sphere.  The luminosity, which is the flux times the surface area, is 
\begin{equation}
\label{eqn:HIIFFLuminosity}
L_{\nu} = 4 \pi^2 R_S^2 B_{\nu} (T_e) \left[1 + \frac{1}{2 \alpha_{\nu}^2 R_S^2} (e^{-2 \alpha_{\nu} R_S} (2 \alpha_{\nu} R_S + 1) - 1)\right].
\end{equation}}

\section{Distribution of $Q_{\rm ion}^{\star}$ for a stellar population}
\label{sec:PopIndivDist}
{\bfnop
To calculate the distribution function of $Q_{\rm ion}^{\star}$, I ran Starburst99 (version 6.0.4) models to find the number of each type of star in a stellar population \citep{Leitherer99}.  I considered a population formed with a continuous star-formation rate of $1.0\ \Msun\ \yr^{-1}$ over ages of 5 Myr and older.  The IMF was assumed to be Salpeter ($dN/dM \propto M^{-2.35}$) over an interval $0.1\ \Msun$ to $100\ \Msun$.  I assumed near Solar metallicity ($Z = 0.02$).  Table~\ref{table:QStellarPopulation} lists the number of each type of ionizing star given these assumptions.  Overall, I find that the number of stars of each type varies little for populations with ages over 10 Myr.  

I took the ionizing photon luminosities of O and early B stars from \citet*{Vacca96}.  As in Table 8 of \citet{Vacca96}, I group subgiants with dwarfs and bright giants with giants; in any case, $Q_{\rm ion}^{\star}$ varies slowly with luminosity class at a given spectral type.  \citet{Vacca96} does not give a $Q_{\rm ion}^{\star}$ for O3.5 stars, so I just take the average values of O3 and O4 stars.  

For Wolf-Rayet stars, I note that in \citet*{Smith02}, late WN stars have $Q_{\rm ion}^{\star} \approx 10^{49.0}\ \sec^{-1}$, early WN stars have $Q_{\rm ion}^{\star} \approx 10^{49.3}\ \sec^{-1}$, late WC stars have $Q_{\rm ion}^{\star} \approx 10^{49.0}\ \sec^{-1}$, and early WC stars have $Q_{\rm ion}^{\star} \approx 10^{49.2}\ \sec^{-1}$.  Finally, for WO stars, I just take $Q_{\rm ion}^{\star} \approx 10^{49.0}\ \sec^{-1}$; there are very few WO stars (about 1 in 10000 stars of all ionizing types), so the exact value should not affect results much.  The ionizing luminosities are also listed in Table~\ref{table:QStellarPopulation}.}

\begin{table*}
\begin{minipage}{170mm}
\caption{Ionizing luminosities in a stellar population with a continuous SFR}
\label{table:QStellarPopulation}
\begin{tabular}{lcccc}
\hline
Type        & $\log_{10} (Q_{\rm ion}^{\star} / \sec^{-1})$ $^a$  & \multicolumn{3}{c}{Number of stars $^b$}\\
            &                                                & 5 Myr & 10 Myr & $\ge 15\ \Myr$\\
\hline
O3.0 V+IV   & 49.87 & 27.81 + 27.65 &  27.81 + 27.65 &  27.81 + 27.65\\
O3.5 V+IV   & 49.79 & 69.70 + 27.89 &  69.70 + 27.89 &  69.70 + 27.89\\
O4.0 V+IV   & 49.70 & 89.49 + 17.83 &  90.07 + 17.83 &  90.07 + 17.83\\
O4.5 V+IV   & 49.61 & 119.9 + 50.22 &  120.1 + 50.22 &  120.1 + 50.22\\
O5.0 V+IV   & 49.53 & 167.9 + 69.52 &  167.9 + 69.52 &  167.9 + 69.52\\
O5.5 V+IV   & 49.43 & 213.5 + 92.90 &  213.5 + 92.90 &  213.5 + 92.90\\
O6.0 V+IV   & 49.34 & 296.4 + 78.71 &  296.4 + 78.71 &  296.4 + 78.71\\
O6.5 V+IV   & 49.23 & 384.8 + 124.8 &  384.8 + 124.8 &  384.8 + 124.8\\
O7.5 V+IV   & 49.00 & 534.4 + 114.8 &  534.4 + 114.8 &  534.4 + 114.8\\
O8.0 V+IV   & 48.87 & 672.2 + 200.0 &  672.2 + 200.0 &  672.2 + 200.0\\
O8.5 V+IV   & 48.72 & 969.8 + 268.2 &  995.7 + 320.6 &  995.7 + 320.6\\
O9.0 V+IV   & 48.56 & 1060  + 235.2 &  1261  + 408.0 &  1261  + 408.0\\
O9.5 V+IV   & 48.38 & 1139  + 237.3 &  1658  + 619.9 &  1658  + 619.9\\
B0.0 V+IV   & 48.16 & 1288  + 2.591 &  2568  + 546.7 &  2661  + 546.7\\
B0.5 V+IV   & 47.90 & 2650  + 0.000 &  4703  + 860.4 &  5516  + 866.6\\
\hline
O3.0 III+II & 49.99 & 25.61 + 7.941 &  25.61 + 7.941 &  25.61 + 7.941\\
O3.5 III+II & 49.92 & 7.300 + 8.495 &  7.300 + 8.495 &  7.300 + 8.495\\
O4.0 III+II & 49.86 & 13.03 + 8.649 &  13.03 + 8.649 &  13.03 + 8.649\\
O4.5 III+II & 49.80 & 22.84 + 12.22 &  22.84 + 12.22 &  22.84 + 12.22\\
O5.0 III+II & 49.73 & 28.14 + 13.92 &  28.14 + 13.92 &  28.14 + 13.92\\
O5.5 III+II & 49.65 & 29.29 + 19.56 &  29.30 + 19.56 &  29.30 + 19.56\\
O6.0 III+II & 49.58 & 43.43 + 23.31 &  43.43 + 23.31 &  43.43 + 23.31\\
O6.5 III+II & 49.50 & 51.97 + 32.81 &  51.97 + 32.82 &  51.97 + 32.82\\
O7.5 III+II & 49.32 & 68.17 + 42.49 &  68.17 + 42.49 &  68.17 + 42.49\\
O8.0 III+II & 49.22 & 84.84 + 52.96 &  84.84 + 52.96 &  84.84 + 52.96\\
O8.5 III+II & 49.12 & 136.3 + 79.38 &  136.3 + 79.38 &  136.3 + 79.38\\
O9.0 III+II & 48.97 & 140.6 + 79.29 &  194.5 + 79.29 &  194.5 + 79.29\\
O9.5 III+II & 48.78 & 97.69 + 68.23 &  193.3 + 68.23 &  193.3 + 68.23\\
B0.0 III+II & 48.55 & 0.000 + 70.95 &  217.6 + 98.38 &  217.6 + 98.38\\
B0.5 III+II & 48.27 & 0.000 + 15.04 &  439.7 + 176.7 &  439.9 + 176.7\\
\hline
O3.0 I      & 50.11 & 236.6 &  236.6 &  236.6\\
O3.5 I      & 50.07 & 44.99 &  44.99 &  44.99\\
O4.0 I      & 50.02 & 41.35 &  41.35 &  41.35\\
O4.5 I      & 49.98 & 45.33 &  45.33 &  45.33\\
O5.0 I      & 49.93 & 45.19 &  45.19 &  45.19\\
O5.5 I      & 49.87 & 53.35 &  53.35 &  53.35\\
O6.0 I      & 49.81 & 42.80 &  42.80 &  42.80\\
O6.5 I      & 49.75 & 49.70 &  49.70 &  49.70\\
O7.5 I      & 49.62 & 59.38 &  59.38 &  59.38\\
O8.0 I      & 49.54 & 22.96 &  22.96 &  22.96\\
O8.5 I      & 49.45 & 43.62 &  43.64 &  43.64\\
O9.0 I      & 49.33 & 12.57 &  12.57 &  12.57\\
O9.5 I      & 49.17 & 39.45 &  39.45 &  39.45\\
\hline
WN Late     & 49.00 & 209.0 &  281.3 &  281.3\\
WN Early    & 49.30 & 175.4 &  177.7 &  177.7\\
WC Late     & 49.00 & 76.91 &  94.08 &  94.08\\
WC Early    & 49.20 & 59.43 &  61.22 &  61.22\\
WO          & 49.00 & 3.070 &  3.070 &  3.070\\
\hline
\end{tabular}
\\$^a$: Number of (hydrogen) ionizing photons emitted per star of the given type.  
\\$^b$: Number of stars in a stellar population of the given age (5 Myr, 10 Myr, $\ge 15\ \Myr$), if the star-formation rate has been constant.  It is normalized to a star-formation rate of $1\ \Msun\ \yr^{-1}$.  For dwarfs and subgiants, the entries are sums of the number of dwarfs and subgiants, respectively; likewise, the entries for giants and bright giants are sums of the number of giants and bright giants, respectively.
\end{minipage}
\end{table*}

\section{Summary of the calculation of integrated flux from a starburst region}
\label{sec:IntegratedFluxCalc}
The uniform slab solution provides the intensity along a single sightline, which can be appropriate if the starburst is resolved.  However, we often are interested in the integrated flux of the starburst.  The sightline will have different lengths as it passes through different parts of the starburst.  I {\bfnop summarize} for the reader how much integrated flux remains after free-free absorption for three common geometries: a disc viewed face-on, a disc viewed edge-on, and a sphere.

\subsection{A face-on disc}
The flux observed at Earth is proportional to the luminosity emitted by the starburst into a unit solid angle, $dL / d\Omega_{\rm em} = \int I dA_{\rm proj}$.  For a face-on disc of radius $R_{\rm SB}$ and midplane-to-edge height $h_{\rm SB}$ (edge-to-edge height $2 h_{\rm SB}$), this is simply $dL/d\Omega_{\rm em} = \pi R_{\rm SB}^2 I$, or
\begin{equation}
\frac{dL}{d\Omega_{\rm em}} = \pi R_{\rm SB}^2 \frac{j}{\alpha_{\rm eff}} \left(1 - e^{-2\alpha_{\rm eff} h_{\rm SB}}\right).
\end{equation}
If the starburst had no absorption, it would have $dL/d\Omega_{\rm em} = 2 \pi R_{\rm SB}^2 h_{\rm SB} j$.  Therefore, the ratio of {\bfnop emergent} flux to flux without absorption is
\begin{equation}
\Psi = \frac{1 - e^{-2\alpha_{\rm eff} h_{\rm SB}}}{2 \alpha_{\rm eff} h_{\rm SB}},
\end{equation}
asymptoting to $1 - \alpha_{\rm eff} h_{\rm SB}$ when $\alpha_{\rm eff}$ is very small and the starburst is nearly transparent.  

The covering fraction for this geometry is just
\begin{equation}
f_{\rm cover} = 1 - e^{-2 \alpha_{\rm cover} h_{\rm SB}}.
\end{equation}

\subsection{An edge-on disc}
If the starburst disc is instead observed edge-on, the sightline length $s$ through the disc varies between 0 and $R_{\rm SB}$.  Defining $y$ as the impact parameter from the centre of the disc, so that $y = \sqrt{R_{\rm SB}^2 - s^2/4}$, the emergent flux is related to
\begin{align}
\nonumber \frac{dL}{d\Omega_{\rm em}} & = 4h_{\rm SB} \frac{j}{\alpha_{\rm eff}} \left(R_{\rm SB} - \int_0^{R_{\rm SB}} e^{-2 \alpha_{\rm eff} \sqrt{R_{\rm SB}^2 - y^2}} dy\right)
\end{align}
The ratio of {\bfnop emergent} flux to flux without absorption is
\begin{equation}
\label{eqn:FRatioEdgeOnDisk}
\Psi = \frac{2}{\pi \alpha_{\rm eff} R_{\rm SB}} \left(1 - \int_0^1 e^{-2\alpha_{\rm eff} R_{\rm SB} \sqrt{1 - u^2}} du\right).
\end{equation}
When the disc is nearly transparent, with small $\alpha_{\rm eff}$, the ratio is approximately $1 - 8 \alpha_{\rm eff} R_{\rm SB} / (3 \pi)$.  

The covering fraction is 
\begin{equation}
\label{eqn:FCoverEdgeOnDisk}
f_{\rm cover} = 1 - \int_0^1 e^{-2 \alpha_{\rm cover} R_{\rm SB} \sqrt{1 - u^2}} du.
\end{equation}
When $\alpha_{\rm cover} R_{\rm SB} \ll 1$, $f_{\rm cover} \approx (\pi / 2) R_{\rm SB} \alpha_{\rm cover}$.  

\subsection{A sphere}
Suppose instead a starburst is a sphere with radius $R_{\rm SB}$.  The emergent flux is now proportional to
\begin{align}
\frac{dL}{d\Omega_{\rm em}} & =  \frac{\pi j}{\alpha_{\rm eff}} \left[R_{\rm SB}^2 + \left(\frac{R_{\rm SB}}{\alpha_{\rm eff}} + \frac{1}{2 \alpha_{\rm eff}^2}\right) e^{-2 \alpha_{\rm eff} R_{\rm SB}} - \frac{1}{2 \alpha_{\rm eff}^2}\right]
\end{align}
In a transparent sphere, the luminosity per solid angle is $dL/d\Omega_{\rm em} = (4/3) \pi R_{\rm SB}^3 j$.  Thus the ratio of {\bfnop emergent} flux to unabsorbed flux is
\begin{equation}
\Psi = \frac{3}{4 \alpha_{\rm eff} R_{\rm SB}^3} \left[R_{\rm SB}^2 + \left(\frac{R_{\rm SB}}{\alpha_{\rm eff}} + \frac{1}{2 \alpha_{\rm eff}^2}\right) e^{-2 \alpha_{\rm eff} R_{\rm SB}} - \frac{1}{2 \alpha_{\rm eff}^2}\right].
\end{equation}
In the limit when $\alpha_{\rm eff}$ is small, this ratio is $1 - (3/4) \alpha_{\rm eff} R_{\rm SB}$.  

The covering fraction is
\begin{equation}
f_{\rm cover} = 1 + \frac{1}{2 R_{\rm SB}^2 \alpha_{\rm cover}^2} [e^{-2 \alpha_{\rm cover} R_{\rm SB}} (2 \alpha_{\rm cover} R_{\rm SB} + 1) - 1].
\end{equation}
This reduces to $f_{\rm cover} \approx (4/3) \alpha_{\rm cover} R$ when $\alpha_{\rm cover} R_{\rm SB} \ll 1$.

\end{appendix}

\end{document}